\documentclass[twocolumn,showpacs,preprintnumbers,amsmath,amssymb]{revtex4}
\usepackage{graphicx}
\usepackage{dcolumn}
\usepackage{bm}
\usepackage{epsfig}

\begin{document}
\title{Diffusion-controlled coalescence, fragmentation and collapse of $d$-dimensional $A$-particle islands in the $B$-particle sea}
\author{ Boris ~M.~Shipilevsky}

\affiliation {Institute of Solid State Physics, Chernogolovka,
Moscow district, 142432, Russia\\}

\date{\today}

\begin{abstract}
We present a systematic analysis of diffusion-controlled evolution
and collapse of two identical spatially separated $d$-dimensional
$A$-particle islands in the $B$-particle sea at propagation of the
sharp reaction front $A+B\to 0$ at equal species diffusivities. We
show that at a sufficiently large initial distance between the
centers of islands $2\ell$ compared to their characteristic
initial size and a relatively large initial ratio of
concentrations island/sea the evolution dynamics of the
island-sea-island system is determined unambiguously by the
dimensionless parameter $\Lambda={\cal N}_{0}/{\cal N}_{\Omega}$,
where ${\cal N}_{0}$ is the initial particle number in the island
and ${\cal N}_{\Omega}$ is the initial number of sea particles in
the volume ${\Omega}=(2\ell)^{d}$. It is established that a) there
is a $d$-dependent critical value $\Lambda_{\star}$ above which
island coalescence occurs; b) regardless of $d$ the centers of
each of the islands move towards each other along a {\it
universal} trajectory merging in a united center at the
$d$-dependent critical value $\Lambda_{s}\geq\Lambda_{\star}$; c)
in one-dimensional systems $\Lambda_{\star}=\Lambda_{s}$,
therefore at $\Lambda<\Lambda_{\star}$ each of the islands dies
individually, whereas at $\Lambda>\Lambda_{\star}$ coalescence is
completed by collapse of a single-centered island in the system
center; d) in two- and three-dimensional systems in the range
$\Lambda_{\star}< \Lambda < \Lambda_{s}$ coalescence is
accompanied by subsequent fragmentation of a two-centered island
and is completed by individual collapse of each of the islands. We
discuss a detailed picture of coalescence, fragmentation and
collapse of islands focusing on evolution of their shape and on
behavior of the relative width of the reaction front at the final
collapse stage and in the vicinity of starting coalescence and
fragmentation points. We demonstrate that in a wide range of
parameters the front remains sharp up to a narrow vicinity of the
coalescence, fragmentation and collapse points.
\end{abstract}
]                                                 

\maketitle

\narrowtext

\section{Introduction}

The fundamental reaction-diffusion system $A+B\to 0$, where unlike
species $A$ and $B$ diffuse and irreversibly annihilate in the
bulk of a $d$-dimensional medium, has attracted great interest in
recent decades owing to the remarkable property of {\it effective
dynamical repulsion} of unlike species \cite{kbr}-\cite{cot}. In
unbounded systems with initially statistically homogeneous
particle distribution, this property brings about spontaneous
growth of $A$ and $B$ particles domains (Ovchinnikov-Zeldovich
segregation) and, as a consequence, anomalous reaction
deceleration. In systems with initially spatially separated
reactants this property results in the formation and self-similar
propagation of a localized reaction front which, depending on the
interpretation of $A$ and $B$ (chemical reagents, quasiparticles,
topological defects, etc), plays a key role in a broad spectrum of
problems in physics, chemistry, biology, and materials science
\cite{but1}-\cite{must}.

The simplest model of a planar reaction front, introduced by Galfi
and Racz (GR)\cite{gal} is the quasi-one-dimensional model for two
initially separated reactants which are uniformly distributed on
the left side ($x<0$) and on the right side ($x>0$) of the initial
boundary. Taking the reaction rate in the mean-field form
$R(x,t)=ka(x,t)b(x,t)$ ($k$ being the reaction constant), GR
discovered that in the long time limit $kt\gg 1$ the reaction
profile $R(x,t)$ acquires the universal scaling form
$$
R=R_{f}{\cal R}\left(\frac{x-x_{f}}{w}\right),
$$
where $x_{f}$, $R_{f}$ and $w$ are, respectively, position, height
and width of the reaction front and the front width anomalously
slowly grows with time by the law
$$
w\propto (t/k^{2})^{1/6},
$$
so that on the diffusion length scale $\propto t^{1/2}$ the
relative width of the front asymptotically contracts unlimitedly
$\propto (kt)^{-1/3}\to 0$ as $kt\to\infty$. Subsequently, it was
shown \cite{cor1}-\cite{bar} that the mean-field approximation is
valid at $d>d_{c}=2$, whereas in one-dimensional systems
fluctuations play the dominant role. Nevertheless, the
self-similar front motion takes place at all dimensions so that at
any $d$ on the diffusion length scale the relative front width
contracts asymptotically. Based on this fact a general concept of
the front dynamics for nonzero diffusivities, the quasistatic
approximation (QSA), was developed \cite{cor1}, \cite{lee},
\cite{bar}, \cite{ben}. The key property of the QSA is that the
front width $w(J)$ depends on $t$ only through the time dependent
boundary current, $J_{A}=|J_{B}|=J$, the calculation of which is
reduced to solving the external diffusion problem with the moving
{\it absorbing boundary} (Stefan problem)
$$
R=J\delta(x-x_{f}).
$$
On the basis of the QSA a general description of spatiotemporal
behavior of the system $A+B\to 0$ has been obtained for arbitrary
nonzero diffusivities \cite{koza} which was then generalized to
the cases of anomalous diffusion \cite{yus},\cite{sok}, diffusion
in disordered systems \cite{kzt}, \cite{hec}, diffusion in systems
with inhomogeneous initial conditions \cite{rkz}, and to several
more complex reactions. Following this approach, in most
subsequent works the use of the QSA was traditionally restricted
by the quasi-one-dimensional sea-sea problem with $A$ and $B$
domains having an unlimited extension, i.e. with an unlimited
number of $A$ and $B$ particles, where asymptotically the stage of
monotonous quasistatic front propagation is always reached.

Recently, a new line in the study of the $A+B\to 0$ front dynamics
has attracted significant attention under the assumption that the
particle number of one or both species is {\it finite} (island-sea
and island-island systems) and, therefore, in the final state one
or both islands {\it disappear} completely
\cite{self1}-\cite{self7}. It has been demonstrated that in the
sharp-front regime these systems exhibit rich scaling behavior,
and though in these systems the QSA is always asymptotically
violated, at large initial particle numbers and a high reaction
constant the vast majority of particles die in the {\it
sharp-front} regime over a wide parameter range. Here we will
focus mainly on the island-sea system, introduced originally in
the Ref.\cite{self1} for quasi-one-dimensional geometry (flat
front) and generalized for $d$ dimensions (ring-shaped or
spherical fronts) in the recent Ref. \cite{self7}. This system is
a basic model for a wide range of phenomena and is realized in
numerous applications from Liesegang patterns formation
\cite{fial}-\cite{man} to electron-hole luminescence in quantum
wells \cite{but1}-\cite{but2} depending on the conditions of
initial island formation.

Two related island-sea problems were considered at equal species
diffusivities in Ref.\cite{self7}: (i) the evolution and collapse
of an initially uniform $d$-dimensional spherical $A$-particle
island "submerged" into the uniform $d$-dimensional $B$-particle
sea and (ii) the formation of a $d$-dimensional spherical
$A$-particle island from a localized $A$-particle source acting a
finite time in the $d$-dimensional initially uniform $B$-particle
sea and subsequent evolution and collapse of the island after
source switching-off in the long-living island regime when the
island collapse time $t_{c}$ exceeds significantly the duration of
source action.  It has been established that at sufficiently large
starting number of $A$ particles ${\cal N}_{0}$ (where ${\cal
N}_{0}$ is the initial number of $A$ particles in the initially
uniform island or the number of injected $A$ particles at the
moment of source switching-off) and a sufficiently large reaction
constant $k$ the death of majority of island particles ${\cal
N}(t)$, regardless of the initial particle distribution, proceeds
in the universal scaling regime,
$$
{\cal N}={\cal N}_{0}{\cal G}(t/t_{c})
$$
where $t_{c}\propto {\cal N}_{0}^{2/d}$ is the lifetime of the
island in the sharp-front limit and on the final stage of collapse
$$
{\cal N}/{\cal N}_{0}\propto {\cal T}^{(d+2)/d}\to 0
$$ as ${\cal T}=(t_{c}-t)/t_{c}\to 0$. It has been shown that at a relatively
large starting ratio of concentrations island/sea, regardless of
the starting particle number and the system dimension, while
dying, the island first expands to a certain maximal amplitude and
then begins to contract by the universal law
\begin{eqnarray}
\nonumber
\zeta_{f}=r_{f}/r_{f}^{M}=\sqrt{e\tau|\ln\tau}|,
\end{eqnarray}
where $\tau=t/t_{c}$ and $r_{f}^{M}\propto {\cal N}_{0}^{1/d}$ is
the island maximal expansion radius at the front turning point
$$
t_{M}=t_{c}/e.
$$
According to Ref. \cite{self7} regardless of the system dimension
the evolution of the boundary current density $J$ that determines
the quasistatic front width $w(J)$ is described by the universal
law
$$
{\cal J}=J/J_{M}=\sqrt{\frac{|\ln \tau|}{e\tau}},
$$
whence it follows that in the mean-field regime the relative front
width $\eta=w/r_{f}$ changes by the law
\begin{eqnarray}
\nonumber
\eta=\eta_{M}/(e\tau\ln^{2}\tau)^{1/3},
\end{eqnarray} where at
the front turning point $\eta_{M}\propto 1/{\cal
N}_{0}^{2/3d}k^{1/3}$ and, therefore, on the final stage of
collapse
$$
\eta\sim \left(\frac{{\cal T}_{Q}}{{\cal T}}\right)^{2/3},
$$
where ${\cal T}_{Q}\propto 1/{\cal N}_{0}^{1/d}\sqrt{k}\to 0$ as
${\cal N}_{0}, k\to\infty$. In Ref. \cite{self7}, an exhaustive
analysis of the reaction front relative width evolution for the
fluctuation, the logarithmically modified, and the mean-field
regimes was presented, and it was demonstrated that in a wide
range of parameters at a large enough number of injected or
initially uniformly distributed particles the front remains sharp
up to a narrow vicinity of the island collapse point, and
therefore the whole picture of the evolution and collapse of the
island is completely {\it self-consistent}.

According to Ref. \cite{self7}, with an increase of the initial
particle number in the island the amplitude of island expansion at
the front turning point increase unlimitedly, and, therefore, in
the presence of {\it neighboring islands} in the sea \cite{man},
\cite{but3} the scenario described above for the autonomous
evolution of the island is realized only as long as the amplitude
of the island expansion remains much less than the distance
between the centers of neighboring islands. If in the sea there
are one or several neighboring islands and this condition is
violated, i.e. the amplitude of the island's expansion becomes
comparable with the distance between the centers of the
neighboring islands, it is obvious that the dynamics of island
evolution must radically change.

In this article, for the first time we pose and systematically
investigate the problem of diffusion-controlled {\it interaction}
of two identical $d$-dimensional $A$-particle islands separated by
a sufficiently large initial distance in the $d$-dimensional
$B$-particle sea. This model is the simplest basic model of the
{\it island-sea-island} system which allows revealing the key
features of the evolution dynamics under the assumption of sharp
front formation at equal species diffusivities. Moreover, because
of mirror symmetry, this model simultaneously describes the
evolution of the $d$-dimensional $A$-particle island in a
semi-infinite $B$-particle sea with a reflecting
$(d-1)$-dimensional "wall". We discover that if the initial
distance between the centers of the islands $2\ell$ is large
enough compared to their characteristic initial size and the
initial ratio of concentrations island/sea is relatively large,
the evolution dynamics of the island-sea-island system
demonstrates remarkable universality and is determined
unambiguously by the dimensionless parameter
\begin{eqnarray}
\nonumber
\Lambda={\cal N}_{0}/{\cal N}_{\Omega},
\end{eqnarray}
where ${\cal N}_{0}$ is the initial particle number in the island
and ${\cal N}_{\Omega}$ is the initial number of sea particles in
the volume ${\Omega}=(2\ell)^{d}$. We show that at
$\Lambda^{2/d}\ll 1$ each of the islands evolves and dies {\it
autonomously} not feeling the presence of a neighboring island and
demonstrate that there is a $d$-dependent critical value
$\Lambda_{\star}$ below which each of the islands dies
individually and above which island coalescence occurs. We also
reveal that there is the second $d$-dependent critical value
$\Lambda_{s}\geq\Lambda_{\star}$ above which coalescence is
completed by collapse of the formed single-centered island in the
system center and we discover the remarkable fact that at $d\geq
2$ in the range $\Lambda_{\star}< \Lambda < \Lambda_{s}$
coalescence is accompanied by subsequent fragmentation of the
two-centered island and is completed by individual collapse of
each of the islands. We discuss a detailed picture of coalescence,
fragmentation and collapse of the islands, reveal the remarkable
properties of universality and self-similarity of the evolution of
islands, give a comprehensive picture of the relative front width
evolution, and demonstrate that in a wide range of parameters the
reaction front remains sharp up to a narrow vicinity of the
coalescence, fragmentation and collapse points.

\section{Evolution of two identical spatially separated
$d$-dimensional $A$-particle islands in the $d$-dimensional
$B$-particle sea}

\subsection{Model}

We consider a model in which two identical $A$-particle islands,
which for simplicity have the shape of a $d$-dimensional hypercube
with the side $2h$ and the centers of which are located on the $x$
axis at the points $x=\pm \ell$, are surrounded by a uniform
unlimited $B$-particle sea with the initial concentration $b_{0}$.
We shall assume that initially in each of the islands
$A$-particles are distributed uniformly with the concentration
$a_{0}$. We shall also assume that initially the islands have the
same spatial orientation and that the coordinate axes with the
origin at the point $x=0$ on the $x$ axis are normal to hypercube
"faces" so that full symmetry takes place
$$
x\leftrightarrow -x, y\leftrightarrow -y, z\leftrightarrow -z.
$$
Particles $A$ and $B$ diffuse with the diffusion constants
$D_{A,B}$, and when meeting they annihilate with some nonzero
probability, $A+B\to 0$. In the continuum version, this process
can be described by the reaction-diffusion equations
\begin{eqnarray}
\partial a/\partial t = D_{A}\nabla^{2} a - R, \quad
\partial b/\partial t = D_{B}\nabla^{2} b - R,
\end{eqnarray}
where $a({\bf r},t)$ and $b({\bf r},t)$ are the mean local
concentrations of $A$ and $B$ and $R({\bf r},t)$ is the
macroscopic reaction rate. We shall assume, as usual, that species
diffusivities are equal $D_{A}=D_{B}=D$. This important condition,
due to local conservation of difference concentration $a-b$, leads
to a radical simplification that permits to obtain an analytical
solution for arbitrary front trajectory (at different species
diffusivities $D_{A}\neq D_{B}$ an analytical solution of the
Stefan problem is possible only for stationary or a monotonically
moving front \cite{self5}). Then, by measuring the length, time
and concentration in units of $h, h^{2}/D$, and $b_{0}$,
respectively, and defining the ratio $a_{0}/b_{0}=c$ and the ratio
$L=\ell/h\gg 1$, we come from Eq. (1) to the simple diffusion
equation for the difference concentration $s({\bf r},t)= a({\bf
r},t)- b({\bf r},t)$,
\begin{eqnarray}
\partial s/\partial t = \nabla^{2} s,
\end{eqnarray}
at the initial conditions
\begin{eqnarray}
s_{0}(|x|\in (L-1,L+1))=c,
\end{eqnarray}
and $s_{0}=-1$ (sea) outside the islands in the 1D case,
\begin{eqnarray}
s_{0}(|x|\in (L-1,L+1), y\in (-1,+1))=c,
\end{eqnarray}
and $s_{0}=-1$ (sea) outside the islands in the 2D case,
\begin{eqnarray}
s_{0}(|x|\in (L-1,L+1), y,z\in (-1,+1))=c,
\end{eqnarray}
and $s_{0}=-1$ (sea) outside the islands in the 3D case, with the
boundary conditions
\begin{eqnarray}
s(|{\bf r}|\to\infty,t)=-1
\end{eqnarray}
and the symmetry conditions
$$
\partial_{x} s\mid_{x=0}=\partial_{y}
s\mid_{y=0}=\partial_{z} s\mid_{z=0}=0.
$$

\subsection{Universal long-time asymptotics in the sharp-front limit}

Exact solution of the problem Eqs. (2)-(6) has the form
\begin{eqnarray}
s(x,t)+1= \frac{(c+1)}{2}({\cal L}_{+}+{\cal L}_{-})
\end{eqnarray}
in the 1D case,
\begin{eqnarray}
s({\bf r},t)+1=\frac{(c+1)}{2}({\cal L}_{+}+{\cal L}_{-})Q(y,t)
\end{eqnarray}
in the 2D case, and
\begin{eqnarray}
s({\bf r},t)+1=\frac{(c+1)}{2}({\cal L}_{+}+{\cal
L}_{-})Q(y,t)Q(z,t)
\end{eqnarray}
in the 3D case, where
\begin{eqnarray}
{\cal L}_{+}(x,t)={\rm erf}\left(\frac{L+1+x}{2\sqrt{t}}\right)-
{\rm erf}\left(\frac{L-1+x}{2\sqrt{t}}\right),
\end{eqnarray}
\begin{eqnarray}
{\cal L}_{-}(x,t)={\rm erf}\left(\frac{L+1-x}{2\sqrt{t}}\right)-
{\rm erf}\left(\frac{L-1-x}{2\sqrt{t}}\right),
\end{eqnarray}
and
\begin{eqnarray}
Q(v,t)= \frac{1}{2}\left[{\rm
erf}\left(\frac{1+v}{2\sqrt{t}}\right)+{\rm
erf}\left(\frac{1-v}{2\sqrt{t}}\right)\right].
\end{eqnarray}
As well as in Refs.\cite{self1},\cite{self7}, we shall assume that
the ratio of concentrations island/sea is large enough, $c\gg 1$
(concentrated island). Below it will be shown that in the limit of
large $c\gg 1$ the "lifetime" of the islands $t_{c}\gg 1$, so the
majority of the $A$-particles die at times $t\gg 1$, when the
diffusive length exceeds appreciably the initial island size. The
evolution of the islands in such a large-$t$ regime is of
principal interest for us here since, as will be demonstrated
below, in the limit of large $t\gg 1$, $L\gg 1$ and $c\gg 1$ {\it
regardless of the initial shape, orientation and sizes} of the
islands the asymptotics of island evolution takes a {\it
universal} form which at a given initial sea density is determined
unambiguously only by the initial number of particles in the
islands ({\it the instantaneous source regime}) and the initial
distance between their centers.

Assuming that the diffusion length $\sqrt{t}\gg 1$ and expanding
the functions ${\cal L}_{+}(x,t)$, ${\cal L}_{-}(x,t)$ and
$Q(v,t)$ in powers of $1/\sqrt{t}$ we find
\begin{eqnarray}
{\cal L}_{+}(x,t)=\frac{2e^{-(L+x)^{2}/4t}}{\sqrt{\pi
t}}(1+q_{+}),
\end{eqnarray}
\begin{eqnarray}
{\cal L}_{-}(x,t)=\frac{2e^{-(L-x)^{2}/4t}}{\sqrt{\pi
t}}(1+q_{-}),
\end{eqnarray}
\begin{eqnarray}
Q(v,t)=\frac{e^{-v^{2}/4t}}{\sqrt{\pi
t}}\left(1-\frac{(1-v^{2}/2t)}{12 t}+\cdots\right).
\end{eqnarray}
where
$$
q_{\pm}=\frac{1}{12t}\left[\frac{(L\pm
x)^{2}}{2t}-1\right]+\cdots,
$$
and the terms of a higher order of smallness in powers of $1/t$,
$(L\pm x)^{2}/t^{2}$ and $v^{2}/t^{2}$, respectively, are omitted
(following the leading term in $q_{\pm}$ has the form
$$
\frac{1}{160t^{2}}\left[1-\frac{(L\pm x)^{2}}{t} +\frac{(L\pm
x)^{4}}{12t^{2}}\right]).
$$
According to the QSA in the diffusion-controlled limit at large
$k\to\infty$ at times $t\propto k^{-1}\to 0$, there forms a sharp
reaction front $w/|{\bf r}_{f}|\to 0$ so that in neglect of the
reaction front width the solution $s({\bf r},t)$ defines the the
law of its propagation
$$
s({\bf r}_{f},t)=0
$$
and the evolution of particles distributions $a({\bf r},t)=s({\bf
r},t)>0$ within the island and $b({\bf r},t)=|s({\bf r},t)<0|$
beyond it. Considering the domain $x>=0$ in view of
$x\leftrightarrow -x$ symmetry and assuming that $|x-L|\ll t,L$,
from Eqs. (13), (14) we conclude that at $1\ll t\ll L^{2}$, when
the diffusion length is much less than the initial distance
between the island centers, the ratio ${\cal L}_{+}/{\cal L}_{-}$
is exponentially small (${\cal L}_{+}/{\cal L}_{-}\sim t
e^{-Lx/t}/L$ at $1\ll t \ll L$ and $\sim e^{-xL/t}$ at $L\ll t \ll
L^{2}$) Therefore, neglecting the contribution of ${\cal L}_{+}$
and assuming that the radius of a $d$-dimensional sphere with the
center at the point of initial island center $\rho\ll t$, we find
from Eqs.(7)-(9)
$$
s(\rho,t)+1=\frac{(c+1)e^{-\rho^{2}/4t}}{(\pi
t)^{d/2}}(1-\xi_{d}),
$$
where
$$
\xi_{d}=\frac{(d-\rho^{2}/2t)}{12t}+\cdots.
$$
Neglecting further the term $\xi_{d}\ll 1$, we conclude that in
agreement with Ref.\cite{self7} regardless of the initial island
shape (hypercube or hypersphere) at $1\ll t\ll L^{2}$ each of the
islands takes the shape of a $d$-dimensional sphere with the front
radius $\rho_{f}(t)$ which changes by the law
\begin{eqnarray}
\rho_{f}(t)=\sqrt{2dt\ln(t_{c}/t)},
\end{eqnarray}
whence it follows that at any $d$ in the limit of large $c\gg 1$
the island first expands reaching some maximal radius
$\rho_{f}^{M}$, and then it contracts disappearing in the collapse
point
\begin{eqnarray}
t_{c}=\frac{(c+1)^{2/d}}{\pi}=\frac{(\gamma N_{0})^{2/d}}{4\pi}
\end{eqnarray}
where $\gamma=(c+1)/c\approx 1$, $N_{0}$ is the initial particle
number in the island in units of $h^{d}b_{0}$ and at the front
turning point $t_{M}=t_{c}/e$
\begin{eqnarray}
\rho_{f}^{M}=\sqrt{2dt_{M}}=(\gamma N_{0})^{1/d}\sqrt{d/2\pi e}.
\end{eqnarray}
At large $t_{c}$ the requirement $\chi_{d}^{f}\ll 1$ along with
the requirement $t\gg 1$ obviously reduces to the more rigid
requirement $t\gg \ln(t_{c}/t)$. On the other hand, the
requirement of "autonomous" death of each of the islands $t_{c}\ll
L^{2}$ reduces to the requirement
\begin{eqnarray}
\Lambda^{2/d}\ll 1, \quad \Lambda = (c+1)/L^{d}.
\end{eqnarray}

\section{Evolution of the island-sea-island system in the
instantaneous source regime}

According to Eqs. (13)-(16), at large $L\gg 1$ in the domain $t\gg
{\sf Max}[1,\ln(t_{c}/t)]$ evolution of the island bounded by the
front becomes independent on its initial size, therefore the
initial distance between the island centers $2\ell$ becomes the
only length scale determining the evolution. Then by measuring the
length and time in units of $\ell$ and $\ell^{2}/D$, i.e. going to
the dimensionless variables $T=t/L^{2}, X=x/L, Y=y/L, Z=z/L$, and
neglecting the transient terms $q_{\pm},(v/t)^{2}\ll 1$ in Eqs.
(13)-(15) we find from Eqs. (7)-(9) and (13)-(15)
\begin{eqnarray}
s+1=\frac{2\Lambda}{(\pi
T)^{d/2}}\exp\left(-\frac{1+X^{2}+\varrho^{2}}{4T}\right)\cosh\left(\frac{X}{2T}\right)
\end{eqnarray}
where $\varrho^{2}=Y^{2}$ or $\varrho^{2}=Y^{2}+Z^{2}$ at $d=2,3$,
respectively. Taking $s_{f}=0$ we derive from Eq. (20) the law of
the reaction front motion
\begin{eqnarray}
\exp\left(-\frac{1+X_{f}^{2}+\varrho_{f}^{2}}{4T}\right)\cosh\left(\frac{X_{f}}{2T}\right)=\frac{(\pi
T)^{d/2}}{2\Lambda}
\end{eqnarray}
and we conclude that in the instantaneous source regime evolution
dynamics of the island-sea-island system is determined
unambiguously by the value of the parameter $\Lambda$ which in
view of the requirement $c\gg 1 (\gamma\approx 1)$ is the ratio of
the initial particle number in the island ${\cal N}_{0}$ to the
initial number of sea particles ${\cal N}_{\Omega}$ in the volume
$\Omega=(2\ell)^{d}$
$$
\Lambda = {\cal N}_{0}/{\cal N}_{\Omega}.
$$
From Eq.(20) it follows that at any $d$ the points where the
concentration of $A$-particles reaches its maximum, which
according to Refs. \cite{self3}, \cite{self4} we will call the
{\it island centers}, are located on the $X$ axis. Calculating the
trajectories of motion of the centers $X_{\star}(T)$ from the
condition $\partial s/\partial X=0$, we obtain from Eq.(20)
\begin{eqnarray}
\tanh \left(\frac{X_{\star}}{2T}\right)=X_{\star},
\end{eqnarray}
whence at small $T\ll 1$ we find
$$
|X_{\star}|=1-2e^{-X_{\star}/2T}+\cdots, \quad |X_{\star}|/T\gg 1,
$$
whereas at small $|X_{\star}|/T\ll 1$ we have
$$
|X_{\star}|=\sqrt{6(T_{s}-T)}+\dots, \quad |X_{\star}|/T\ll 1,
$$
where $T_{s}=1/2$. We conclude thus that with increasing $T$,
regardless of the dimension of the system and the value of the
parameter $\Lambda$, the centers of both islands move towards each
other along the universal trajectory (22) from $X_{\star}=\pm 1$
to $|X_{\star}|\to 0$, merging at $T_{s}=1/2$ into the single
center $X_{\star}=0$ in the system center ${\bf r}={\bf 0}$. It is
clear, however, that mutual convergence of the island centers
caused by effective diffusion-controlled "attraction" of the
islands continues only till the collapse moment $T_{c}(\Lambda)$
of each of the islands which depends on the quantity $\Lambda$. It
is also clear that after expansion and subsequent contraction of
the islands collapse of each of them is completed at the point of
the corresponding center
$$
X_{c}=X_{\star}(T_{c})
$$
the coordinates of which are fixed by the system of equations
which follows from Eqs. (21), (22)
\begin{eqnarray}
\tanh \left(\frac{X_{c}}{2T_{c}}\right)=X_{c},
\end{eqnarray}
\begin{eqnarray}
\exp\left(-\frac{1+X_{c}^{2}}{4T_{c}}\right)\cosh\left(\frac{X_{c}}{2T_{c}}\right)=\frac{(\pi
T_{c})^{d/2}}{2\Lambda}.
\end{eqnarray}
According to Eq. (21), at $T\leq T_{s}$ on the $X$ axis each of
the islands is bounded by two leading front points (by two fronts
in the 1D case) $|X_{f}^{-}|< |X_{\star}|$ and
$|X_{f}^{+}|>|X_{\star}|$ which determine the width of the island
$$
|X_{f}^{-}|<|X_{\sf isl}|<|X_{f}^{+}|
$$
and, therefore, the time moment of the island collapse is
determined by the condition
$$
|X_{f}^{-}(T_{c})|=|X_{f}^{+}(T_{c})|=|X_{\star}(T_{c})|.
$$
With growing $\Lambda$ the distance to the system center
$|X_{f}^{-}|$ at the front turning point obviously reduces till at
some critical value $\Lambda_{\star}$ both of the leading front
points $\pm X_{f}^{-}$ merge in the system center $|X_{f}^{-}|=0$
and, thus, at $\Lambda> \Lambda_{\star}$ coalescence of the
islands occurs with the formation of a united island with the
half-width $|X_{f}^{+}|$. According to Eq. (20), in the system
center we find
\begin{eqnarray}
s({\bf 0},T)+1=\frac{2\Lambda}{(\pi
T)^{d/2}}\exp\left(-\frac{1}{4T}\right)
\end{eqnarray}
whence it follows immediately that $s({\bf 0},T)$ reaches the
maximum $s_{M}({\bf 0})={\sf Max}_{T} [s({\bf 0},T)]$ at the time
moment
$$
T_{M}=1/2d
$$
whence we find
$$
s_{M}({\bf 0})= 2\Lambda \left(\frac{2d}{\pi e}\right)^{d/2} - 1
$$
Assuming further that $s_{M}({\bf 0})=0$ we obtain finally the
critical point of coalescence threshold
\begin{eqnarray}
\Lambda_{\star}=\frac{1}{2}\left(\frac{\pi
e}{2d}\right)^{d/2}=\left\{\begin{array} {lcl} 1.03318..., \quad
d=1,\\
1.06747..., \quad d=2,\\
0.84900..., \quad d=3.\\
\end{array}\right.
\end{eqnarray}
Substituting now $T=T_{s}$ into Eq. (25) and assuming that $s({\bf
0},T_{s})=0$, we find the critical point of threshold of island
centers merging
\begin{eqnarray}
\Lambda_{s}=\frac{\sqrt{e}}{2}\left(\frac{\pi}{2}\right)^{d/2}=\left\{\begin{array}
{lcl} \Lambda_{\star}, \quad
d=1,\\
1.29490..., \quad d=2,\\
1.62292..., \quad d=3.\\
\end{array}\right.
\end{eqnarray}
above which the formed single-centered island dies in the system
center. Indeed, according to Eq. (20), in the system center we have
$$
\partial^{2} s/\partial X^{2}\mid_{{\bf r}={\bf 0}}= - \frac{2\Lambda
e^{-1/4T}}{(\pi T)^{d/2}}(1-T_{s}/T)
$$
whence, according to Eq.(22), it follows that at the critical
point $T_{s}=1/2$ the transition local minimum of $s$
$\rightarrow$ global maximum of $s$ occurs
$$
s({\bf 0},T< T_{s})={\sf Min}_{X}(s)\rightarrow s({\bf
0},T>T_{s})={\sf Max}_{X}(s).
$$
From Eqs.(25)-(27) we conclude that regularities of the
island-sea-island system evolution differ qualitatively at $d=1$
and $d>1$. In 1D systems $\Lambda_{\star}=\Lambda_{s}$, that is
why in the domain $\Lambda < \Lambda_{\star}$ each of the islands
dies individually not touching the partner, whereas above the
threshold $\Lambda > \Lambda_{\star}$ the single-centered island
formed during coalescence dies in the system center. At $d
>1$ in the range $\Lambda_{\star}< \Lambda < \Lambda_{s}$
a united (dumbbell-like) two-centered island is formed which again
splits into two separated islands (fragmentation) at some moment
$T_{fr}(\Lambda)$ with subsequent death in corresponding centers
$\pm X_{\star}(T_{c})$. It is easy to understand the reasons for
absence of the intermediate coalescence-fragmentation domain in 1D
systems. Indeed, in 2D and 3D systems the sea always remains
topologically continuous ({\it pathwise-connected}), that is why
after formation of an isthmus between the islands (coalescence)
the current of sea particles normal to the $X$ axis strives to
destroy the isthmus ($A+B\to 0$) and reach this (fragmentation) in
the range $\Lambda_{\star}< \Lambda< \Lambda_{s}$ as the island is
depleted. In a qualitative contrast to that, in 1D systems the sea
consists of two areas separated by the islands: a finite
"internal" sea area enclosed between the fronts $\pm X_{f}^{-}$
($0\leq |X_{\sf isea}|< |X_{f}^{-}|$) and an unbounded "external"
sea $|X_{f}^{+}|<|X_{\sf sea}|<\infty$.  Thus, after disappearance
of the internal sea area (coalescence) collapse of the formed
island in the system center is the only remaining outcome of the
reaction in 1D systems.

From Eq.(25) it follows that above the coalescence threshold
$\Lambda > \Lambda_{\star}$ the condition $s({\bf 0},T)=0$ leads
to occurrence of two roots $T_{0}^{(-)} < T_{M}$ and $T_{0}^{(+)}>
T_{M}$. The first of these roots, $T_{0}^{(-)}$, determines the
starting time of island coalescence
$$
T_{0}^{(-)}=T_{cl}(\Lambda)
$$
and unlimitedly (logarithmically slowly) decreases with an
increase in $\Lambda$
$$
T_{cl}\propto 1/4\ln\Lambda, \quad \ln\Lambda\gg 1.
$$
The meaning of the second of these roots, $T_{0}^{(+)}$, depends
on system dimension and value of $\Lambda$. In 1D systems
$T_{0}^{(+)}$ determines the collapse point of the formed
single-centered island
$$
T_{0}^{(+)}=T_{c}(\Lambda).
$$
At $d>1$ in the range $T_{M}< T_{0}^{(+)}< T_{s}$
($\Lambda_{\star}< \Lambda < \Lambda_{s}$) the second root gives
the fragmentation point of the two-centered island
$$
T_{0}^{(+)}=T_{fr}(\Lambda),
$$
whereas at $T_{0}^{(+)}>T_{s}$ it determines the collapse time of
the single-centered island which increases unlimitedly with
growing $\Lambda$
$$
T_{c}\propto (2\Lambda)^{2/d}/\pi, \quad \Lambda^{2/d}\gg 1.
$$
It is important to note that in the limit of large
$\Lambda^{2/d}\gg 1$ at $T\gg {\sf Max}[\ln(T_{c}/T),1]$ the term
$\cosh(X_{f}/2T)\approx 1$ in Eq.(21) can be neglected, therefore,
in the course of evolution the island takes the shape of a
$d$-dimensional sphere, the radius of which (just as in the
autonomous evolution domain $\Lambda^{2/d}\ll 1$ (Eq. (16))
changes by the law
\begin{eqnarray}
|X_{f}^{+}|=\sqrt{X_{f}^{2}+\varrho_{f}^{2}}\approx\sqrt{2dT\ln(T_{c}/T)}
\end{eqnarray}
where
$$
T_{c}=\frac{(2\Lambda)^{2/d}}{\pi}(1-c_{d}/\Lambda^{2/d}+\cdots)\approx
(2\Lambda)^{2/d}/\pi
$$
(with $c_{d}=\pi/2^{(d+2)/d}d$) instead of
$T_{c}=(\Lambda^{2/d})/\pi$ in the domain of autonomous evolution.
This result is a trivial consequence of the fact that in the limit
of large $T$, when diffusion length becomes much larger than the
initial distance between the islands, the evolution of the island
formed during coalescence should obey asymptotically the law of
evolution from an instantaneous source with the twice initial
number of particles $2{\cal N}_{0}$. According to Eq. (20), in 2D
and 3D systems the half-width (radius) of the isthmus between the
islands in the section $X=0$ grows during coalescence by the law
$$
|\varrho_{0f}|=\sqrt{4T\ln[2\Lambda/(\pi T)^{d/2}]-1}
$$
reaching the maximum
$$
|\varrho_{0f}^{M}|=\sqrt{(\Lambda/\Lambda_{\star})^{2/d}-1}
$$
at the time moment
$$
T_{M}^{\varrho}=(\Lambda/\Lambda_{\star})^{2/d}/2d.
$$
whence in the upper limit of the fragmentation domain
$\Lambda=\Lambda_{s}$ we obtain
$$
T_{M}^{\varrho}(\Lambda_{s})/T_{s}=e^{(1-d)/d}=\left\{\begin{array}
{lcl} 0.60653..., \quad
d=2,\\
0.51341..., \quad d=3,\\
\end{array}\right.
$$
and
$$
|\varrho_{0f}^{M}(\Lambda_{s})|=\sqrt{de^{(1-d)/d}-1}=\left\{\begin{array}
{lcl} 0.46158..., \quad
d=2,\\
0.73501..., \quad d=3.\\
\end{array}\right.
$$
Correspondingly, in the limit of large $\Lambda^{2/d}\gg 1$ under
expansion and subsequent contraction of the $d$-dimensional sphere
in accord with Eq. (28) we find
$T_{M}^{\varrho}(\Lambda\to\infty)/T_{c}=1/e$ and
$|\varrho_{0f}^{M}(\Lambda\to\infty)|=(\Lambda/\Lambda_{\star})^{1/d}$
at the front turning point. Fig. 1 demonstrates the dependencies
$T_{c}(\Lambda), T_{cl}(\Lambda)$ and $T_{fr}(\Lambda)$ calculated
from Eqs. (23-25) for $d=1,2,3$. These dependencies reveal a
comprehensive picture of location and extension of the domains of
autonomous death of the islands (I), individual death of the
islands below the coalescence threshold (II),
coalescence-fragmentation of the two-centered island with
subsequent individual death of each of the islands (III) and
coalescence-collapse of the single-centered island in the system
center (IV).

\begin{figure}
\includegraphics[width=0.9\columnwidth]{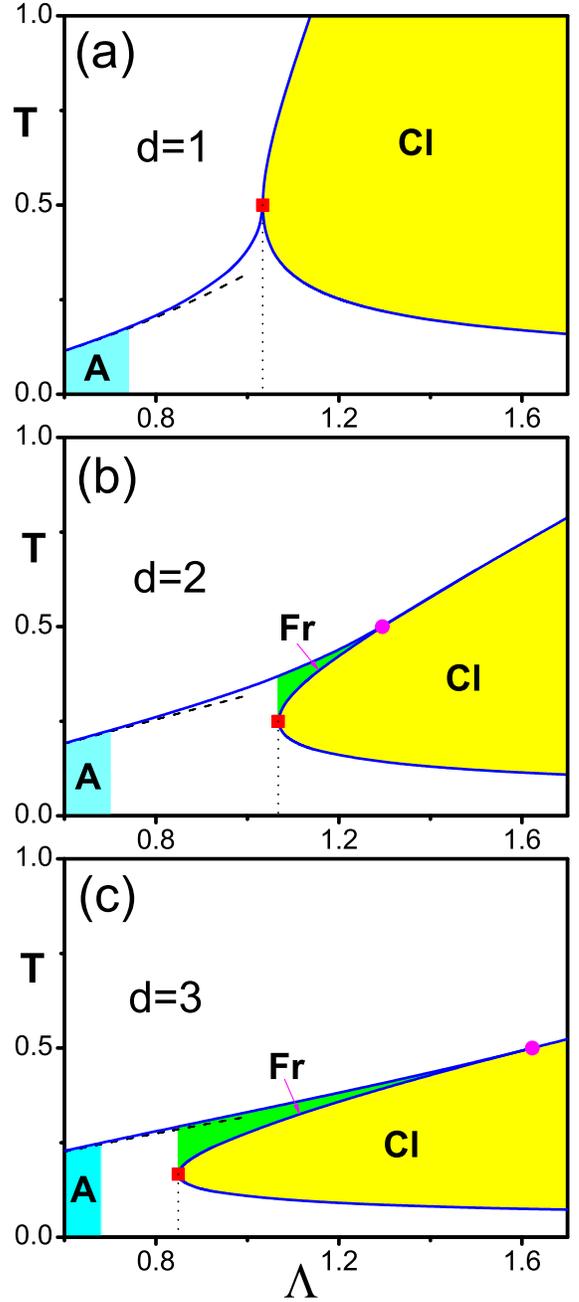}
\caption {Dependencies $T_{c}(\Lambda), T_{cl}(\Lambda)$ and
$T_{fr}(\Lambda)$ calculated from Eqs. (23)-(25) for $d$=1 (a),
$d$=2 (b) and $d$=3 (c). The areas of autonomous collapse,
coalescence and fragmentation are colored. The critical points
$(\Lambda_{\star}, T_{M})$ and $(\Lambda_{s}, T_{s})$ are marked
by square and circle, respectively. The dashed lines show the
asymptotics of the autonomous collapse [Eq.(16)].}
\label{Fig 1}
\end{figure}

\section{Evolution of front trajectories}

Figures 2,3 and 4 show the dependencies $|X_{f}^{-}(T)|,
|X_{f}^{+}(T)|$ and $|X_{\star}(T)|$ calculated from Eqs. (7)-(12)
($L=10^{2}, v=0$) and Eqs. (21),(22) ($\varrho=0$) for $d=1,2$ and
$3$, respectively. These dependencies, in combination with the
behavior of the velocities along trajectories, demonstrate the key
features of evolution of trajectories of fronts and of island
centers with growing parameter $\Lambda$.

\subsection{1D systems}

From Fig.2 it is seen that in 1D systems with increasing $\Lambda$
the mutual "self-accelerating" convergence of the island centers
is accompanied by the corresponding asymmetric "deformation" of
the front trajectories.

A) The trajectory $|X_{f}^{-}(T)|$. In the domain $\Lambda<
\Lambda_{a}^{-}\approx 0.982\Lambda_{\star}$ the velocity of front
motion $|V_{f}^{-}|=|dX_{f}^{-}/dT|$ along the trajectory
$|X_{f}^{-}|$ decreases monotonically up to the front turning
point $V_{f}^{-}=0$ after the passage of which both of the fronts
$|X_{f}^{\pm}|$ move towards each other accelerating up to the
point of island collapse $|V_{f}^{\pm}(T\to T_{c})|\to\infty$. At
$\Lambda_{a}^{-}<\Lambda <\Lambda_{\star}$ on the trajectory
$|X_{f}^{-}(T)|$ two inflection points, ${\sf Min}|V_{f}^{-}|$ and
${\sf Max}|V_{f}^{-}|$, arise between which the domain of front
acceleration appears and expands. As $\Lambda$ approaches
$\Lambda_{\star}$, the amplitude of ${\sf Max}|V_{f}^{-}|$
increases unlimitedly ${\sf Max}|V_{f}^{-}|(\Lambda\to
\Lambda_{\star})\to\infty$ decreasing abruptly to $0$ at the front
turning point with subsequent rapid collapse of the island. The
reason for this behavior is obviously the competition between two
opposing trends: (i) the striving of each of the islands to
"destroy" the finite (internal) sea area as motion of the front
$|X_{f}^{-}|$ accelerates and (ii) the striving of the unlimited
sea area to "destroy" the island as motion of the front
$|X_{f}^{+}|$ accelerates. At the stage of accelerating motion of
the front $|X_{f}^{-}|$ the process (i) is dominant, whereas after
passage of the point ${\sf Max}|V_{f}^{-}|$ the process (ii) wins
the competition. It is remarkable that precisely at the critical
point $\Lambda_{\star}$ both of the islands and the internal sea
area die {\it simultaneously} at the time moment $T=T_{s}$:
$|X_{f}^{\pm}|\to 0, |V_{f}^{\pm}|\to\infty$ as $T\to T_{s}$. In
the domain of coalescence $\Lambda > \Lambda_{\star}$ the process
(i) wins the competition, that is why the front velocity grows
unlimitedly up to the coalescence point: $|X_{f}^{-}|\to 0,
|V_{f}^{-}|\to\infty$ as $T\to T_{cl}$.

B) The trajectory $|X_{f}^{+}(T)|$. In the domain of individual
island collapse $\Lambda< \Lambda_{\star}$ after passage of the
front turning point $V_{f}^{+}=0$ the front velocity monotonically
increases unlimitedly everywhere up to the point of island
collapse $T_{c}$: $|V_{f}^{+}|\to\infty$ as $T\to T_{c}< T_{s}$.
At $\Lambda > \Lambda_{\star}$ two inflection points, ${\sf
Max}|V_{f}^{+}|$ and ${\sf Min}|V_{f}^{+}|$, arise on the
trajectory $|X_{f}^{+}|$ between which the domain of front motion
deceleration appears and expands. With an increase in $\Lambda$
the amplitude of ratio ${\sf Max}|V_{f}^{+}|/{\sf Min}|V_{f}^{+}|$
decrease rapidly, so at large $\Lambda\gg \Lambda_{\star}$ the
domain of front deceleration actually disappears. Since the domain
of front deceleration arises in the vicinity $T\approx T_{s}<
T_{c}$, it is qualitatively clear that the reason for front motion
deceleration is a rapid increase in the concentration of island
particles in the system center against a background of merging of
the island centers in a united center at $T= T_{s}=T_{M}$ where
the concentration reaches the maximum. According to Fig.2, with
increasing $\Lambda$, by the time moment $T\approx T_{s}$ the
distance from the front to the system center rapidly increases
and, as a consequence, the effect of passage through the maximum
in the system center on front motion decreases up to complete
disappearance at large $\Lambda$.

\begin{figure}
\includegraphics[width=1\columnwidth]{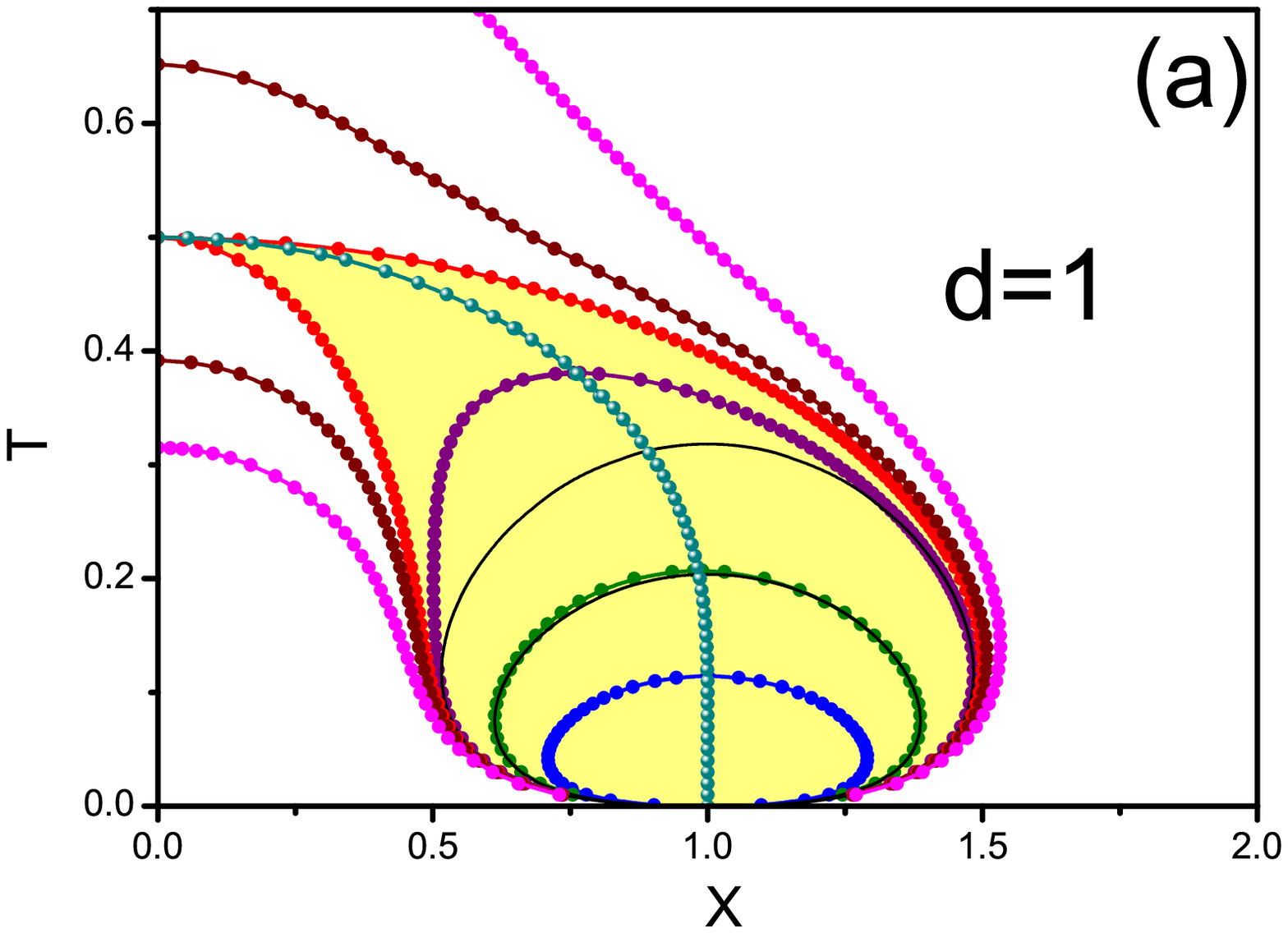}
\label{Fig2a}
\end{figure}
\begin{figure}
\includegraphics[width=0.5\columnwidth]{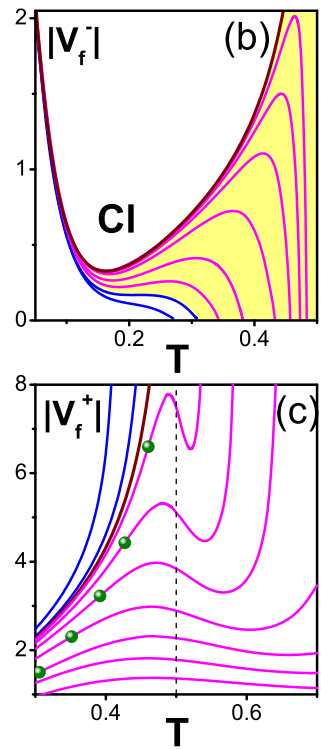}
\caption {1D systems: (a) The trajectories $|X_{f}^{+}(T)|$ and
$|X_{f}^{-}(T)|$ calculated for $\Lambda=0.6, 0.8, 1,
\Lambda_{\star}=1.03318, 1.05$ and $1.1$ according to Eqs.
(7)-(12) $(L=10^{2}, v=0)$ (filled circles) and Eq. (21)
$(\varrho=0)$ (thick lines). The area of individual collapse below
the coalescence threshold is colored. Semi-filled circles show the
trajectory of the island center $|X_{\star}(T)|$ (Eq. (22). The
trajectories of autonomous collapse calculated from Eq. (16) for
$\Lambda=0.6, 0.8$ and $1$ are shown by thin lines; (b) Time
dependencies $|V_{f}^{-}(T)|$ up to the turning point calculated
from Eq. (21) for $\Lambda=1.01, 1.015, 1.02, 1.025, 1.03, 1.032,
1.0325, 1.033$ and $\Lambda_{\star}$ (from bottom to top). The
area with two inflection points is colored; (c) Time dependencies
$|V_{f}^{+}(T)|$ calculated from Eq. (21) for $\Lambda=1.02, 1.03,
\Lambda_{\star}, 1.035, 1.04, 1.05, 1.07, 1.11, 1.14$ and $1.2$
(from top to bottom). The circles mark the starting points of
coalescence.}
\label{Fig2}
\end{figure}

\subsection{2D and 3D systems}

From Fig.3 it is seen that in the 2D case, as well as in the 1D
case, at $\Lambda > \Lambda_{a}^{-}\approx 0.974\Lambda_{\star}$
two inflection points, ${\sf Min}|V_{f}^{-}|$ and ${\sf
Max}|V_{f}^{-}|$, arises on the trajectory $|X_{f}^{-}(T)|$
between which the domain of front acceleration appears and
expands. In a qualitative contrast to 1D systems, however, as
$\Lambda$ approaches $\Lambda_{\star}$ the amplitude of ${\sf
Max}|V_{f}^{-}|$ approaches the finite value $|V_{f\star}^{-}|$
which is reached precisely at the critical point
$\Lambda=\Lambda_{\star}$ where $|X_{f}^{-}|\to 0, |V_{f}^{-}|\to
|V_{f\star}^{-}|$ as $T\to T_{M}$. Moreover, at
$\Lambda=\Lambda_{\star}$ at the time moment $T=T_{M}$ "elastic"
reflection of the front from the system center occurs with a
sudden reversal of velocity sign:
$$
\pm |V_{f}^{-}|(T=T_{M}-0)\rightarrow \mp |V_{f}^{-}|(T=T_{M}+0).
$$
In the coalescence-fragmentation domain $\Lambda_{\star}< \Lambda
<\Lambda_{s}$ the front $|X_{f}^{-}|$, moving to the system
center, disappears at the coalescence point ($|X_{f}^{-}|\to 0,
|V_{f}^{-}|\to \infty$ as $T\to T_{cl}<T_{M}$), and arises again
at the fragmentation point ($|X_{f}^{-}|\to 0, |V_{f}^{-}|\to
\infty$ as $T\to T_{fr}<T_{s}$), moving to the island center, with
its subsequent collapse at the point $T_{c}$. In a qualitative
contrast to 1D systems, after passage of the turning point
$|V_{f}^{+}|=0$ the front $|X_{f}^{+}|$ moves with an unlimitedly
increasing velocity at any $\Lambda$ up to the collapse point
$|X_{f}^{+}|\to |X_{c}|, |V_{f}^{+}|\to\infty$ as $T\to T_{c}$. 3D
systems demonstrate the similar behavior (Fig.4) with
$\Lambda_{a}^{-}\approx 0.957\Lambda_{\star}$.

\begin{figure}
\includegraphics[width=1\columnwidth]{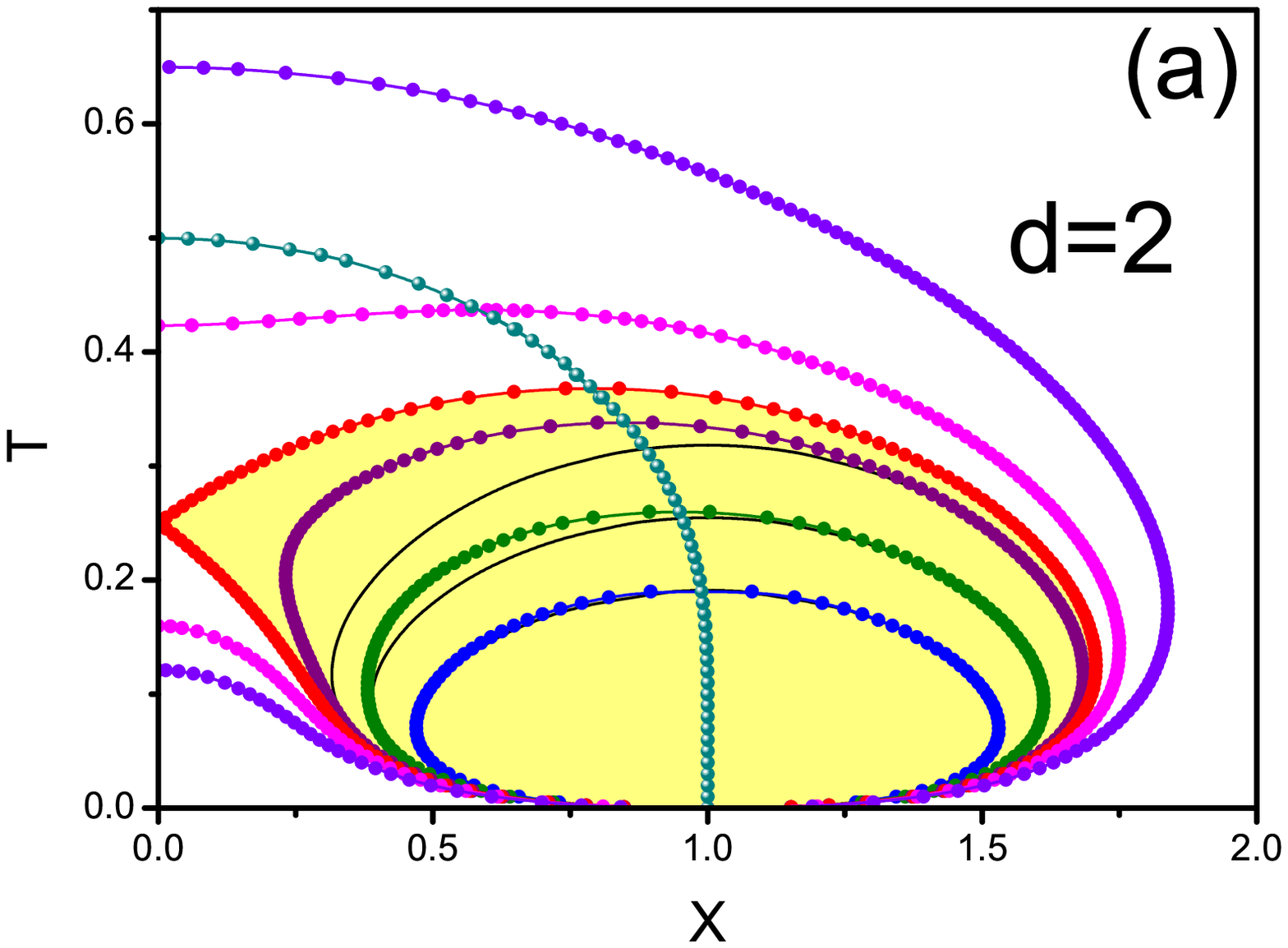}
\label{Fig3a}
\end{figure}
\begin{figure}
\includegraphics[width=0.5\columnwidth]{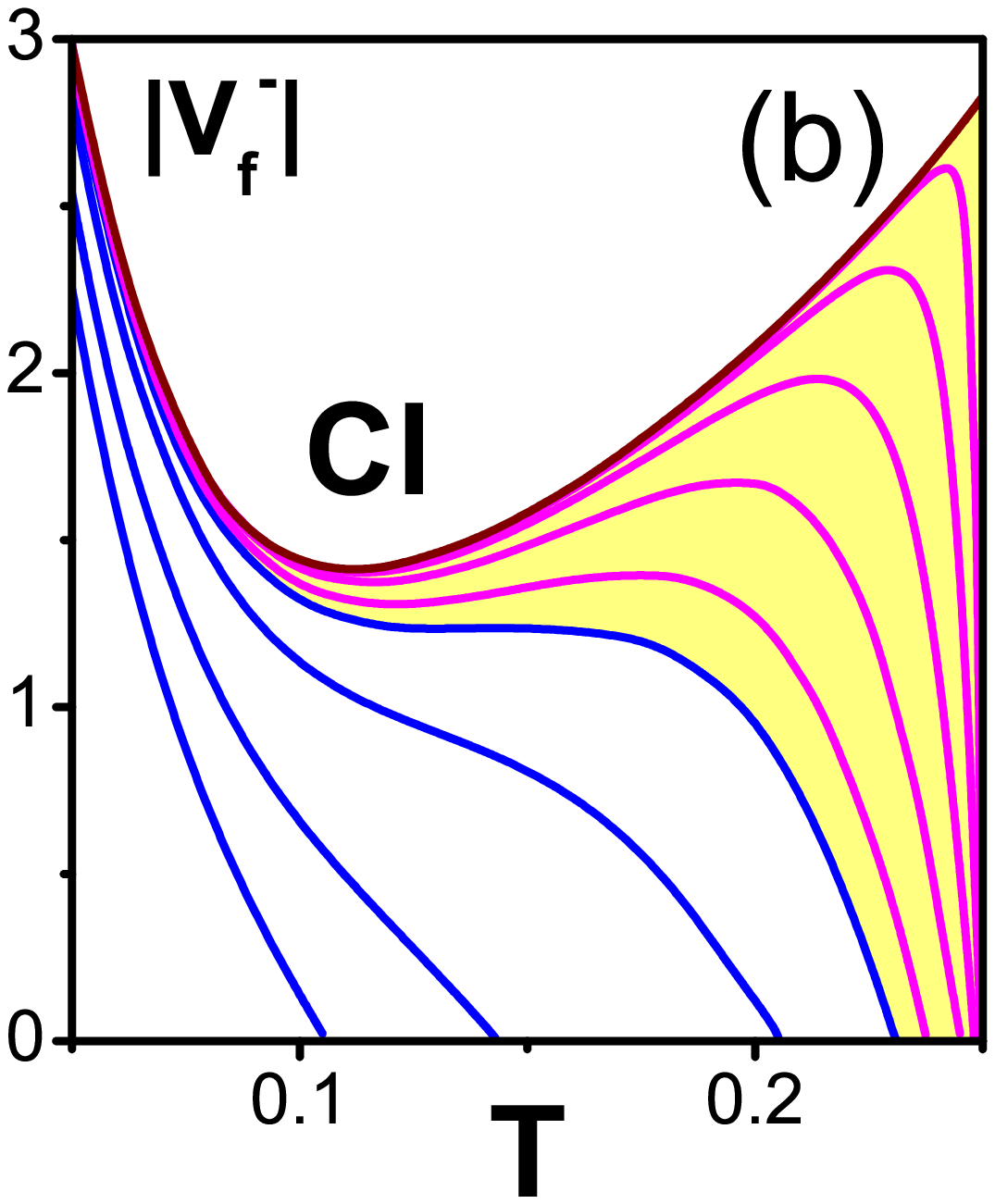}
\caption {2D systems: (a)The trajectories $|X_{f}^{+}(T)|$ and
$|X_{f}^{-}(T)|$ calculated for $\Lambda=0.6, 0.8, 1,
\Lambda_{\star}=1.06747, 1.2$ and $1.5$ according to Eqs. (7)-(12)
$(L=10^{2}, v=0)$ (filled circles) and Eq. (21) $(\varrho=0)$
(thick lines). The area of individual collapse below the
coalescence threshold is colored. Semi-filled circles show the
trajectory of the island center $|X_{\star}(T)|$ (Eq. (22). The
trajectories of autonomous collapse calculated from Eq. (16) for
$\Lambda=0.6, 0.8$ and $1$ are shown by thin lines; (b) Time
dependencies $|V_{f}^{-}(T)|$ up to the turning point calculated
from Eq. (21) for $\Lambda=0.08, 0.09, 1, 1.04, 1.05, 1.06, 1.065,
1.067, 1.06744$ and $\Lambda_{\star}$ (from bottom to top). The
area with two inflection points is colored. }
\label{Fig3}
\end{figure}

\begin{figure}
\includegraphics[width=1\columnwidth]{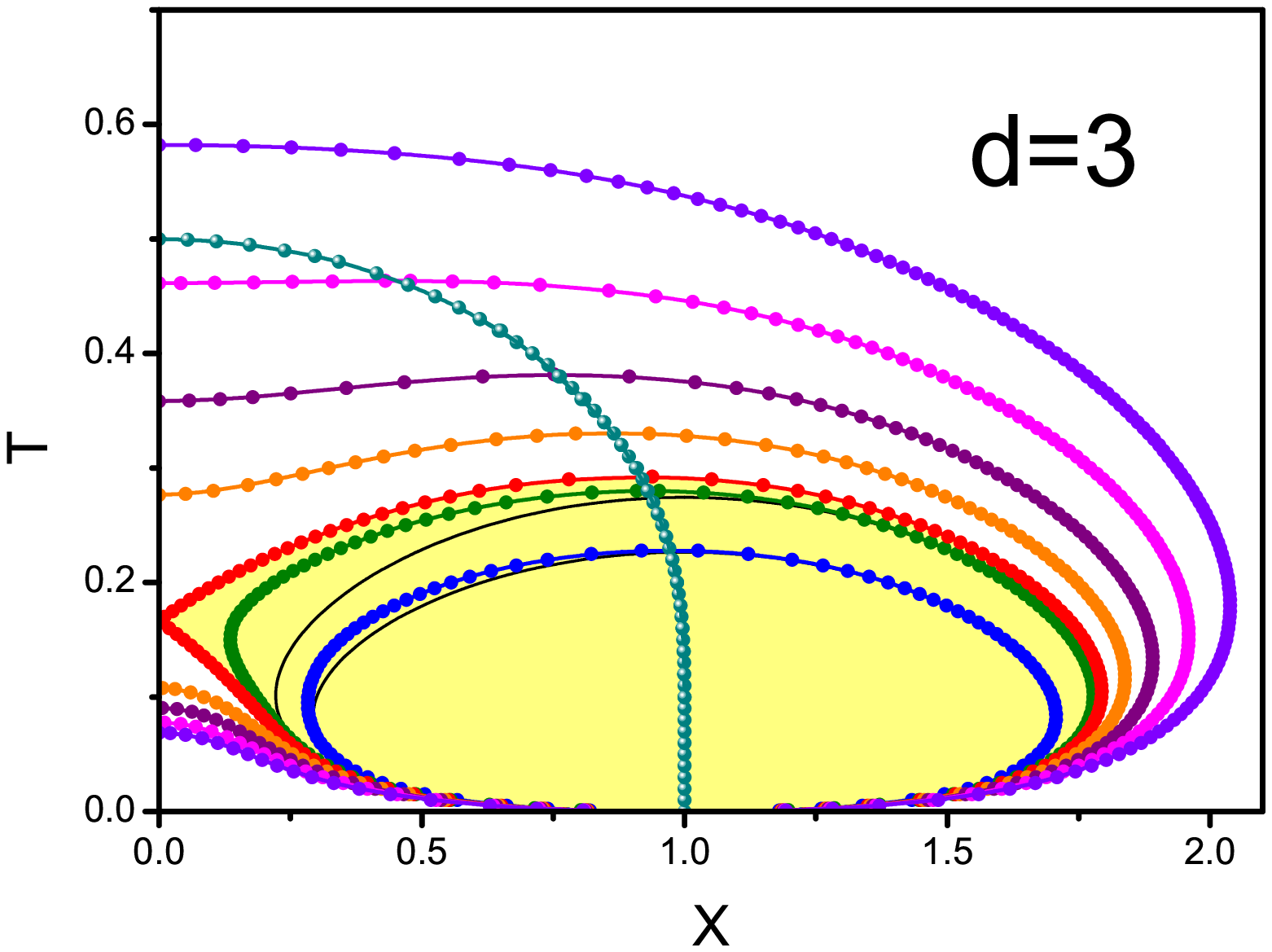}
\caption {3D systems: The trajectories $|X_{f}^{+}(T)|$ and
$|X_{f}^{-}(T)|$ calculated for $\Lambda=0.6, 0.8,
\Lambda_{\star}=0.84900, 1, 1.2, 1.5$ and $1.9$ according to Eqs.
(7)-(12) $(L=10^{2}, v=0)$ (filled circles) and Eq. (21)
$(\varrho=0)$ (thick lines). The area of individual collapse below
the coalescence threshold is colored. Semi-filled circles show the
trajectory of the island center $|X_{\star}(T)|$ (Eq. (22). The
trajectories of autonomous collapse calculated from Eq. (16) for
$\Lambda=0.6$ and $0.8$ are shown by thin lines.}
\label{Fig4}
\end{figure}

\subsection{Front trajectories in the vicinity of coalescence, fragmentation and
collapse points $X_{f}^{\pm}\to 0$}

From Eq. (21) we find that in the limit $X_{f}^{\pm}\to 0$ (in
view of $X\leftrightarrow -X$ symmetry we shall assume that $X\geq
0$)
\begin{eqnarray}
V_{f}X_{f}=\frac{d(1-T_{M}/T+\epsilon_{M})}{(T_{s}/T-1+\epsilon_{s})}
\end{eqnarray}
where
$$
\epsilon_{M}=X_{f}^{2}T_{M}(1/T-1)/T-s({\bf 0},T)+\cdots,
$$
$$
\epsilon_{s}=-X_{f}^{2}(1-1/T+1/12T^{2})/4T^{2}+\cdots,
$$
and $\epsilon_{M,s}\to 0$ as $|T-T_{0}|,X_{f}\to 0$ where
$T_{0}(\Lambda)=T_{0}^{(-)}(\Lambda)=T_{cl}$ or
$T_{0}(\Lambda)=T_{0}^{(+)}(\Lambda)=T_{fr},T_{c}$. Assuming
further that $T_{0}\neq T_{M}, T_{s}$ in the limit $|T-T_{0}|\to
0$ we obtain from Eq. (29)
$$
X_{f}=\sqrt{2d{\cal D}(T_{0}-T)},
$$
and
\begin{eqnarray}
V_{f}=-{\sf sgn}({\cal D})\sqrt{\frac{d{\cal D}}{2(T_{0}-T)}}
\end{eqnarray}
where
$$
{\cal D}=\frac{T_{M}-T_{0}}{T_{s}-T_{0}}.
$$
From Eqs. (30) it follows that at $\Lambda> \Lambda_{\star}$ in 1D
systems ($T_{M}=T_{s}=1/2$) the fronts of collapse
$X_{f}^{+}$$(T_{0}=T_{c}> T_{M})$ and coalescence
$X_{f}^{-}$$(T_{0}=T_{cl}< T_{M})$ reach the system center with
the same reduced velocity {\it regardless} of $\Lambda$
$$
{\cal V}_{f}= |V_{f}|\sqrt{|T_{0}-T|}=1/\sqrt{2}.
$$
In 2D and 3D systems behavior of the fronts changes qualitatively.
With an increase in $\Lambda$ the reduced velocities of
coalescence and fragmentation fronts increase from ${\cal
V}_{f}\to 0$ at $\Lambda\to \Lambda_{\star}+0$ to ${\cal
V}_{f}=1/\sqrt{2}$ at $\Lambda\gg \Lambda_{\star}$ and to ${\cal
V}_{f}\to\infty$ at $\Lambda\to \Lambda_{s}-0$, respectively. In
its turn, with increasing $\Lambda$ the reduced velocity of
collapse of the single-centered island ($\Lambda> \Lambda_{s}$)
decrease from ${\cal V}_{f}\to\infty$ at $\Lambda\to
\Lambda_{s}+0$ to the constant ${\cal V}_{f}=\sqrt{d/2}$ at
$\Lambda\gg \Lambda_{s}$. It is clear that "abnormal" deceleration
${\cal V}_{f}(T_{0}\to T_{M})\to 0$ and acceleration ${\cal
V}_{f}(T_{0}\to T_{s})\to\infty$ of front motion in the vicinity
of the critical points $\Lambda_{\star}$ and $\Lambda_{s}$,
respectively, relate to a radical change in the laws of front
motion at these points. The detailed analysis which will be
presented below shows that at the critical point
$\Lambda=\Lambda_{\star}$ ($T_{0}=T_{M}$) the fronts of
coalescence ($T\to T_{M}-0$) and fragmentation ($T\to T_{M}+0$)
move by the law of "elastic" front reflection
\begin{eqnarray}
X_{f}^{-}=\pm (T_{M}-T)\sqrt{2d^{2}/(d-1)}.
\end{eqnarray}
At the critical point of merging of the centers
$\Lambda=\Lambda_{s}$ we find
\begin{eqnarray}
X_{f}^{+}=[12(d-1)(T_{s}-T)]^{1/4}.
\end{eqnarray}
As we shall see below, first of these results is a direct
consequence of rapid $\propto (T_{M}-T)^{2}$ disappearance of sea
particles in the system center and as a rapid increase in their
concentration after reflection of the front. The second of these
results (which is easily derived from (29)) is a direct
consequence of formation of a superellipse (2D) or superellipsoid
(3D) at the final stage of island collapse. To complete the
picture, we shall also indicate the law of front motion at the
critical point $\Lambda=\Lambda_{\star}=\Lambda_{s}$ for the 1D
case:
\begin{eqnarray}
X_{f}^{\pm}=\sqrt{2(3\pm \sqrt{6})(T_{M}-T)}.
\end{eqnarray}

\section{Evolution of islands in the vicinity of coalescence, fragmentation and collapse points}

In the previous section we focused on front trajectories along the
$X$ axis, $X_{f}^{\pm}(\Lambda,T)|_{\varrho_{f}=0}$, which
determine the evolution of width of the islands and the key
features of their coalescence, fragmentation and collapse. In 1D
systems these trajectories provide comprehensive information on
evolution of islands, whereas in 2D and 3D systems the description
of shape evolution of the islands $\varrho_{f}(\Lambda,T)= {\cal
F}[X_{f}(\Lambda,T)]$ is necessary for a complete picture of their
evolution. In this section, our goal is a detailed analysis of the
evolution of shape of the islands in the vicinity of their
coalescence, fragmentation and collapse points.

Assuming that $|X|/T, X^{2}/T\ll 1$ and $\varrho^{2}/T\ll 1$ we
find from Eq. (20)
\begin{eqnarray}
s+1=\frac{2\Lambda e^{-1/4T}}{(\pi
T)^{d/2}}\left(1-\frac{X^{2}}{4T}{\cal
P}_{2}+\frac{X^{4}}{T^{2}}{\cal P}_{4}-\frac{\varrho^{2}}{4T}
+\cdots\right)
\end{eqnarray}
where
$$
{\cal P}_{2}(T)=1-1/2T
$$
and
$$
{\cal P}_{4}(T)=(1-1/T+1/12T^{2})/32.
$$
Let now as before $s({\bf 0},T_{0})=0$ where, depending on the
value of $\Lambda$, the time moment $T_{0}$ is the point of
coalescence $T_{cl}$, fragmentation $T_{fr}$ or collapse of the
single-centered island $T_{c}$. Then, introducing the reduced time
${\cal T}= (T_{0}-T)/T_{0}$ in the limit of small $|{\cal T}|\ll
1$ we obtain from Eq. (34)
\begin{eqnarray}
s= s({\bf 0},T) - \frac{X^{2}}{4T_{0}}{\cal
P}_{2}^{0}+\frac{X^{4}}{T^{2}}{\cal
P}_{4}-\frac{\varrho^{2}}{4T}+\cdots
\end{eqnarray}
where
$$
{\cal P}_{2}^{0}=(1-T_{s}/T_{0})(1+s({\bf 0},T))+{\cal
T}(1-1/T_{0})+\cdots,
$$
$$
s({\bf 0},T)={\cal T}d(1-T_{M}/T_{0})/2+m{\cal T}^{2}+\cdots,
$$
and
$$
m=\frac{(d+2)}{8}(d-1/T_{0})+1/32T_{0}^{2}.
$$
Assuming further that $s_{f}=0$ we derive from Eq.(35)
\begin{eqnarray}
s({\bf 0},T)=\frac{X_{f}^{2}}{4T_{0}}{\cal
P}_{2}^{0}-\frac{X_{f}^{4}}{T^{2}}{\cal
P}_{4}+\frac{\varrho_{f}^{2}}{4T}+\cdots.
\end{eqnarray}

\subsection{Self-similar evolution of islands at the final collapse stage at $\Lambda\geq\Lambda_{s}$}

\subsubsection{Self-similar collapse at the critical point $\Lambda_{s}=\Lambda_{\star}$ of 1D systems}

At the critical point $\Lambda=\Lambda_{s}$ we have
$T_{0}=T_{s}=T_{c}=1/2$ whence it follows
$$
{\cal P}_{2}^{0}= -{\cal T}+\cdots,
$$
$$
s({\bf 0},{\cal T})= {\cal T}(d-1)/2+(d^{2}-3){\cal
T}^{2}/8+\cdots,
$$
and we derive from Eq.(36)
\begin{eqnarray}
s({\bf 0},{\cal T})= - X_{f}^{2}{\cal
T}/2+X_{f}^{4}/12+\rho_{f}^{2}/2+\dots.
\end{eqnarray}
In the 1D case, where $\Lambda_{s}=\Lambda_{\star}$,$T_{s}=T_{M}$
and $\varrho=0$, from Eq.(37) we reproduce immediately the result
of (33)
$$
X_{f}^{\pm}=\sqrt{(3\pm \sqrt{6}){\cal T}}.
$$
From Eq. (35) a remarkable fact follows that in 1D systems at the
final collapse stage the distribution of particles in the island
and internal area of the sea takes the universal scaling form
\begin{eqnarray}
s(X,{\cal T})={\cal T}^{2}\Phi(|X|/\sqrt{\cal T}).
\end{eqnarray}
As a consequence of this fact, we conclude that at the critical
point $\Lambda_{s}=\Lambda_{\star}$ at the final collapse stage
the ratio of island width $X_{f}^{+}-X_{f}^{-}$ to the half-width
of the internal sea area $X_{f}^{-}$
$$
g_{I/S}=\frac{X_{f}^{+}-X_{f}^{-}}{X_{f}^{-}}=\sqrt{\frac{3+\sqrt{6}}{3-\sqrt{6}}}-1=2.14626
$$
and the ratio of widths front/center $|X_{f}^{\pm}-X_{\star}|$
(where, according to Eq. (22), $X_{\star}=\sqrt{3{\cal T}}$)
$$
g_{\pm}=
\frac{X_{f}^{+}-X_{\star}}{X_{\star}-X_{f}^{-}}=\frac{\sqrt{3+\sqrt{6}}-\sqrt{3}}{\sqrt{3}-\sqrt{3-\sqrt{6}}}=
0.60839...
$$
remain constant up to the collapse point. According to Eq. (38),
in the scaling regime the number of island particles should
decrease by the law ${\cal N}\propto {\cal T}^{5/2}$. The exact
calculation from Eq.(35) gives
$$
{\cal N}/{\cal N}_{0}=m_{1}{\cal T}^{5/2}
$$
where $m_{1}=0.24312...$. Calculating further the ratio of
particle number on the half-width of the internal sea area to
particle number in the island we obtain ${\cal N}_{isea}/{\cal
N}=0.24118...$ whence it follows that at the final collapse stage
the majority of island particles, $\approx 3/4$, die in the
"external" front $X_{f}^{+}$, whereas only $\approx 1/4$ of island
particles die in the "internal" front $X_{f}^{-}$. From Fig.5 it
is seen that the calculated from Eq. (20) normalized particle
distribution $s(X/\sqrt{\cal T})/{\cal T}^{2}$ converges to
scaling function (38) at ${\cal T}\approx < 0.04$ whence we
conclude that $\sim 10^{-4}$ of the initial number of particles
die in the scaling regime (38).

\begin{figure}
\includegraphics[width=1\columnwidth]{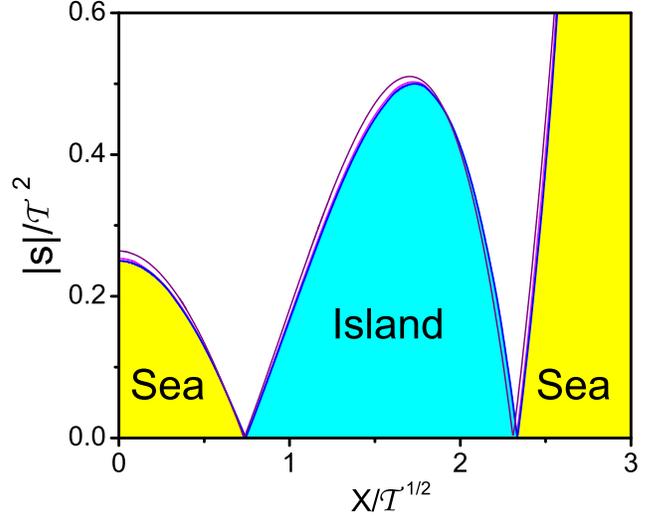}
\caption {Collapse of the normalized distribution of particles
$|s(|X|/\sqrt{{\cal T}})|/{\cal T}^{2}$ to the scaling function
(38) at $\Lambda=\Lambda_{\star}$. Thin lines - ${\cal T}=0.04,
0.01$, thick line - ${\cal T}\to 0$ [Eqs.(20)]. The areas under
scaling function $\Phi(|X|/\sqrt{\cal T})$ are colored.}
\label{Fig5}
\end{figure}

\subsubsection{Final stage of collapse at the critical point $\Lambda_{s}$ of 2D and 3D systems}

In a radical contrast to 1D systems, where at the critical point
of merging of the centers $T_{s}=1/2$ both of the islands
disappear at the moment of island contact
($X_{f}^{+}(T_{s})=X_{f}^{-}(T_{s})=0$), in 2D and 3D systems long
before island collapse a united two-centered island is formed
($T_{cl}< T_{M}< T_{s}$) which disappears at the point of merging
of the centers ($X_{f}^{+}(T_{s})=0$). From Eq.(37) it follows
that at $d>1$, at the final collapse stage ${\cal T}\ll 1$, the
two-centered dumbbell-like island takes the shape of a
superellipse (2D) or superellipsoid (3D)
\begin{eqnarray}
\left(\frac{X_{f}}{X_{f}^{m}}\right)^{4}+\left(\frac{\varrho_{f}}{\varrho_{f}^{m}}\right)^{2}=1
\end{eqnarray}
where, according to Eq. (32), the major semi-axis of the
superellipse (superellipsoid) contracts by the law
$$
X_{f}^{m}({\cal T})=X_{f}^{+}({\cal T})= [6(d-1){\cal T}]^{1/4},
$$
whereas its minor semi-axis contracts by the law
$$
\varrho_{f}^{m}({\cal T})=[(d-1){\cal T}]^{1/2}
$$
and, therefore, the aspect ratio of the superellipse
(superellipsoid) contracts by the law
$$
{\cal A}=\frac{\varrho_{f}^{m}}{X_{f}^{m}}= \left[\frac{(d-1){\cal
T}}{6}\right]^{1/4}\to 0
$$
as ${\cal T}\to 0$. Thus, we conclude that in 2D and 3D systems at
the critical point $\Lambda_{s}$ the island asymptotically takes
the shape of a quasi-one-dimensional "string" the length of which
contracts unlimitedly by the law $\propto {\cal T}^{1/4}$ as
${\cal T}\to 0$. From Eq. (35) it follows that at the final
collapse stage distribution of particles in the island takes the
universal scaling form
\begin{eqnarray}
s={\cal T}{\cal F}_{d}\left(\frac{|X|}{{\cal T}^{1/4}},
\frac{|\varrho|}{{\cal T}^{1/2}}\right),
\end{eqnarray}
whence, taking into account Eq.(39), for the number of particles
in the island we obtain
$$
{\cal N}/2{\cal N}_{0}=m_{d}{\cal T}^{(2d+3)/4},
$$
where $m_{2}= 0.15091...$ and $m_{3}=0.32025...$. Fig. 6 presents
the calculated from Eq. (21) evolution of front at the final stage
of 2D island collapse in the scaling coordinates $Y_{f}/{\cal
T}^{1/2}$ vs. $X_{f}/{\cal T}^{1/4}$. It is seen that the shape of
the island converges to superellips (39) at ${\cal T}<\sim
10^{-3}$. The same picture is observed in the 3D case whence we
conclude that $\sim 10^{-6}$(2D) and $\sim 10^{-7}$(3D) of the
initial number of particles, respectively, die in the scaling
regime (40).

\begin{figure}
\includegraphics[width=1\columnwidth]{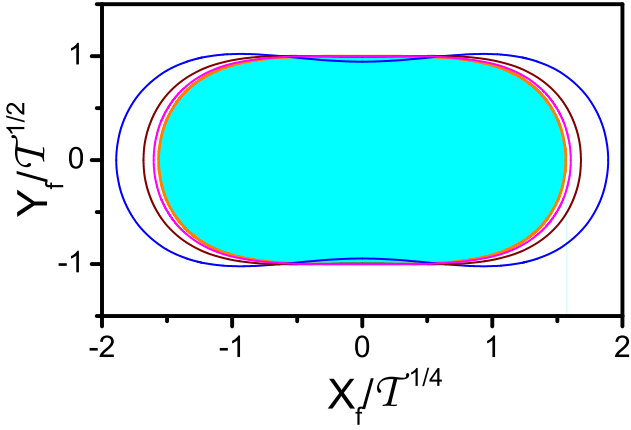}
\caption {Final stage of 2D island evolution at the critical point
$\Lambda=\Lambda_{s}$: collapse of island shape to the
superellipse in the scaling coordinates $Y_{f}/{\cal T}^{1/2}$ vs.
$X_{f}/{\cal T}^{1/4}$. Thin lines - ${\cal T}=0.1, 0.01$ and
$0.001$ (Eq. 21), thick line - Eq. (39). The area of superellipse
is colored.}
\label{Fig6}
\end{figure}

\subsubsection{Final stage of collapse of single-centered island ($\Lambda >\Lambda_{s}$)}

In agreement with Eq. (30), from Eq. (36) at $\chi=T_{s}/T_{c}<1$
in the limit of small ${\cal T}\to 0$ we asymptotically find
\begin{eqnarray}
X_{f}^{+}= \sqrt{\frac{(d-\chi){\cal T}}{\chi(1-\chi)}}(1+\mu{\cal
T}+\cdots)
\end{eqnarray}
where at $1-\chi\ll 1$ coefficient $|\mu|\propto
(d-\chi)/(1-\chi)^{2}$ whence it follows that in 1D systems the
crossover ${\cal T}^{1/2}\rightarrow {\cal T}^{1/2}$ to
asymptotics (41) is realized at ${\cal T}\ll (1-\chi)$, whereas in
â 2D and 3D systems the crossover ${\cal T}^{1/4}\rightarrow {\cal
T}^{1/2}$ to asymptotics (41) is realized at ${\cal T}\ll
(1-\chi)^{2}$. From Eq. (36) we conclude thus that in 2D and 3D
systems at the final collapse stage ${\cal T}\ll (1-\chi)^{2}$ the
single-centered island takes {\it at any} $\Lambda> \Lambda_{s}$
the shape of an ellipse (2D) or ellipsoid of revolution (3D)
\begin{eqnarray} \left(\frac{X_{f}}{X_{f}^{m}}\right)^{2}+
\left(\frac{\varrho_{f}}{\varrho_{f}^{m}}\right)^{2}=1
\end{eqnarray}
where the major semi-axis of the ellipse (ellipsoid) contracts by
the law
$$
X_{f}^{m}=X_{f}^{+}= \sqrt{\frac{(d-\chi){\cal T}}{\chi(1-\chi)}}
$$
whereas its minor semi-axis contracts by the law
$$
\varrho_{f}^{m}=\sqrt{(d-\chi){\cal T}/\chi}
$$
so that asymptotically the island contracts {\it self-similarly}
with the constant aspect ratio
$$
{\cal A}=\varrho_{f}^{m}/X_{f}^{m}=\sqrt{1-\chi}.
$$
As expected, in the limit $\chi\to 1$ ($\Lambda\to\Lambda_{s}$)
the island inherits asymptotically the shape of a quasi-1D
"string", ${\cal A}(\chi\to 1)\to 0$, whereas according to Eq.
(28) in the opposite limit $\chi\ll 1$ ($\Lambda/\Lambda_{s}\gg
1$) the island contracts in the shape of a $d$-dimensional sphere,
${\cal A}(\chi\to 0)\to 1$. According to Eqs.(35) and (42) for
asymptotics of the particle number in the island we find
\begin{eqnarray}
{\cal N}/2{\cal N}_{0}=q_{d}(\Lambda){\cal T}^{(d+2)/2}
\end{eqnarray}
where
$$
q_{d}(\Lambda)=
\frac{\alpha_{d}(d-\chi)^{(d+2)/2}}{\Lambda\chi^{d/2}\sqrt{1-\chi}}
$$
and $\alpha_{1}=1/6$, $\alpha_{2}=\pi/32$, $\alpha_{3}=\pi/60$.
From Eq. (43) it follows that in the vicinity of the critical
point $1-\chi\ll 1$ by the time of crossover to asymptotics
(41)$\propto (1-\chi)^{(2d+3)/2}$ of the initial number of
particles remain in the $d$-dimensional island. With growing
$\Lambda$ the coefficient $q_{1}(\Lambda)$ increases rapidly, and
the coefficients $q_{2,3}(\Lambda)$ decrease rapidly reaching the
values of $q_{1}(\infty)=\sqrt{2/\pi}/3, q_{2}(\infty)=1/2$ and
$q_{3}(\infty)=3\sqrt{6/\pi}/5$ known for a $d$-dimensional sphere
\cite{self7}.

\subsection{Evolution of 2D and 3D islands in the vicinity
of coalescence and fragmentation points ($\Lambda\geq
\Lambda_{\star}$)}

\subsubsection{Shape of islands at the starting points of coalescence $T_{cl}$
and fragmentation $T_{fr}$}

Let now $\chi_{s}=T_{s}/T_{0}> 1$ where  $T_{0}$ is the starting
point of coalescence ($\chi_{s}>d$) or fragmentation
($1<\chi_{s}<d$) of the islands $s({\bf 0},T_{0})=0$. Then,
according to Eq. (35), we find that at the point $T=T_{0}({\cal
T}=0)$ of contact of the islands $X_{f}^{-}(T_{0})=0$ in the
vicinity $X_{f}\ll {\sf min}(\sqrt{\chi_{s}-1}, T_{0})$ of the
system center the front of each of the islands takes the form of
an angle (2D) or cone of revolution (3D) with a vertex in the
system center ${\bf r}={\bf 0}$ and the $\chi_{s}$-dependent value
of opening angle $2\theta$ where
\begin{eqnarray}
\tan\theta= |\varrho_{f}|/X_{f}= \sqrt{\chi_{s}-1}.
\end{eqnarray}
From Eq. (44) it follows that in the coalescence domain
($T_{cl}<T_{M}$) the angle $\theta_{cl}(\Lambda)$ increases from
$\theta_{cl}(\Lambda_{\star})=\pi/4$ (2D) or $\tan^{-1}\sqrt{2}$
(3D) to $\theta_{cl}(\infty)=\pi/2$ with an increase in $\Lambda$,
whereas in the fragmentation domain ($T_{M}<T_{fr}<T_{s}$) the
angle $\theta_{fr}(\Lambda)$ decreases from
$\theta_{fr}(\Lambda_{\star})=\theta_{cl}(\Lambda_{\star})$ to
$\theta_{fr}(\Lambda\to \Lambda_{s})\to 0$ with in increase in
$\Lambda$ . We conclude thus that, as expected, (a) at any
$\Lambda_{\star}< \Lambda < \Lambda_{s}$ the angle of coalescence
is always greater than that of fragmentation:
$$
\theta_{cl}(\Lambda)> \theta_{fr}(\Lambda)
$$
and b) in the limit $\Lambda\to \Lambda_{s} (\chi_{s}\to 1)$ at
the moment of start of fragmentation $T_{fr}$ both of the islands
"inherit" the shape of a quasi-1D "string" $|\varrho_{f}|/X_{f}\to
0$. According to Eq. (35), in this limit the distribution of
particles in each of the islands is determined by the expression
$$
s = (\chi_{s}-1)X^{2}/2 - X^{4}/12 - \varrho^{2}/2+\cdots
$$
whence at the point of fragmentation $T_{fr}$ the island width is
$$
X_{f}^{+}=\sqrt{6(\chi_{s}-1)},
$$
the coordinate of the island center is
$X_{\star}=X_{f}^{+}/\sqrt{2}$, the concentration of $A$-particles
in the island center is $s_{\star}=3(\chi_{s}-1)^{2}/4$, the
amplitude $|\varrho_{f\star}|$ in the island center is
$|\varrho_{f\star}|=\sqrt{3/2}(\chi_{s}-1)$ so that
$|\varrho_{f\star}|/X_{f}^{+}=\sqrt{\chi_{s}-1}/2$ and the
fraction of particles remaining in each of the islands is ${\cal
N}/{\cal N}_{0}\propto (\chi_{s}-1)^{(2d+3)/2}$. As an
illustration, Fig. 7 presents the sequential stages of
coalescence, fragmentation and collapse of 2D islands for
$\Lambda=1.2$, which demonstrate the key features of the evolution
of their shape in the range $\Lambda_{\star}< \Lambda<
\Lambda_{s}$.

\begin{figure}
\includegraphics[width=1\columnwidth]{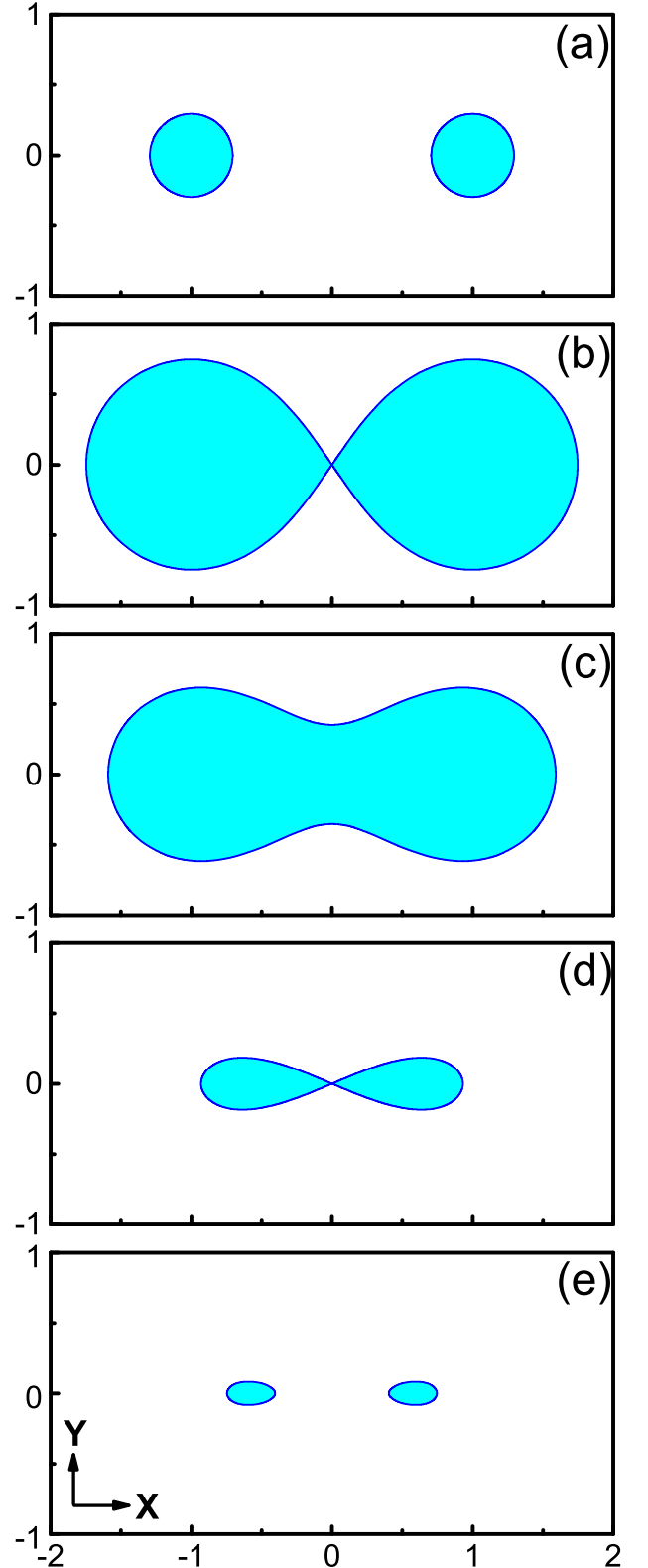}
\caption {Sequential stages of coalescence, fragmentation and
collapse of 2D islands calculated from Eq. (21) at $\Lambda=1.2
(T_{c}=0.4368)$ for the time moments $T=0.005$ (a),
$T=T_{cl}=0.15977$ (b), $T=T_{M}^{\varrho}=0.28104$ (c),
$T=T_{fr}=0.42311$ (d) and $T=0.434$ (e). The areas of islands are
colored.}
\label{Fig7}
\end{figure}

\subsubsection{"Elastic" reflection of the front at the critical point
$\Lambda=\Lambda_{\star}$}

According to Eq. (35), at the critical point
$\Lambda=\Lambda_{\star}$ ($T_{0}=T_{M}$) we find $s({\bf
0},|{\cal T}|)= -d{\cal T}^{2}/4+\cdots$, ${\cal
P}_{2}^{0}=(1-d)+\cdots$ and conclude that at $|{\cal T}|,
X_{f}\ll 1$ the front of each of the islands takes the shape of a
hyperbola (2D) or hyperboloid of revolution (3D)
\begin{eqnarray}
\left(\frac{X_{f}}{X_{f}^{m}}\right)^{2}-\left(\frac{\varrho_{f}}{\varrho_{f}^{m}}\right)^{2}=1
\end{eqnarray}
where real, $X_{f}^{m}=X_{f}^{-}$, and imaginary,
$\varrho_{f}^{m}$, semi-axes of the hyperbola (hyperboloid) first
contract (${\cal T}>0$), and then grow (${\cal T}<0$) by the law
$$
X_{f}^{m}=|{\cal T}|/\sqrt{2(d-1)}, \quad  \varrho_{f}^{m}=|{\cal
T}|/\sqrt{2}
$$
keeping ${\cal T}\leftrightarrow -{\cal T}$ symmetry with a sudden
reversal of the velocity sign at the contact point of vertices of
hyperbola (hyperboloid), ${\cal T}=0$, where, according to Eq.
(44), branches of hyperbola (hyperboloid) degenerate in coaxial
angles (cones) with $\theta(\Lambda_{\star})=\tan^{-1}\sqrt{d-1}$
which determine the asymptotes of the hyperbola (hyperboloid)
$(|\varrho_{f}|/X_{f})_{a}=\sqrt{d-1}$.

\subsubsection{Evolution of islands in the vicinity of coalescence and
fragmentation points $\Lambda>\Lambda_{\star}$}

Assuming $|{\cal T}|\ll {\sf min}[(\chi_{s}-1)^{2}, |\chi_{M}-1|,
T_{0}]$ where $\chi_{M}=\chi_{s}/d=T_{M}/T_{0}$, from Eq. (36) we
find that evolution of island shape in the vicinity of coalescence
and fragmentation points is described by the expression
\begin{eqnarray}
\left(\frac {X_{f}}{X_{f}^{m}}\right)^{2} - \left(\frac
{\varrho_{f}}{\varrho_{f}^{m}}\right)^{2}={\sf
sign}[(\chi_{M}-1){\cal T}]
\end{eqnarray}
where with an increase in $T$ the semi-axes of the hyperbola
(hyperboloid of revolution) first contract (${\cal T}> 0$), and
then grow (${\cal T}< 0$) by the laws
$$
X_{f}^{m}=\sqrt{\frac{d|(\chi_{M}-1){\cal
T}|}{\chi_{s}(\chi_{s}-1)}}
$$
and
$$
\varrho_{f}^{m}=\sqrt{d|(\chi_{M}-1){\cal T}|/\chi_{s}}
$$
with the time-independent asymptotes
$$
(|\varrho_{f}|/X_{f})_{a}= \sqrt{\chi_{s}-1}.
$$
According to Eq. (46), in the coalescence domain ($\chi_{M}>1$)
vertices of the hyperbola (two-sheet or elliptic hyperboloid)
$|X_{f}^{-}|$ move towards each other (${\cal T}>0$),
accelerating, up to the coalescence point ${\cal T}=0$, where the
semi-axis $X_{f}^{m}$ becomes imaginary and $\varrho_{f}^{m}$
(${\cal T}<0$) determines a decelerating increase in width
(radius) of the isthmus of the formed two-centered island
(one-sheet or hyperbolic hyperboloid). In the fragmentation domain
($1/d<\chi_{M}<1$), where the two-centered island divides into two
separated islands, this process occurs in a reverse order:
hyperbolic hyperboloid (${\cal T}>0$)$\rightarrow$ elliptic
hyperboloid (${\cal T}<0$). It is remarkable that in the limit of
a quasi-1D "string" $\chi_{s}-1\ll 1$ Eq. (35) allows describing
explicitly a complete picture of fragmentation up to the point of
individual collapse of each of the islands. For $d>1$ at $|{\cal
T}|, |X|\ll 1$ from Eq. (36) we have
\begin{eqnarray}
{\cal T}(d-1)=-X_{f}^{2}(\chi_{s}-1+{\cal
T})+X_{f}^{4}/6+\varrho_{f}^{2}+\cdots,
\end{eqnarray}
whence for the trajectories $X_{f}^{\pm}({\cal T})$ of the front
points along the $X$ axis ($\varrho_{f}=0$) we find
$$
(X_{f}^{\pm})^{2}=3(\chi_{s}-1+{\cal T})\pm
\sqrt{9(\chi_{s}-1+{\cal T})^{2}+6{\cal T}(d-1)}
$$
and, as a consequence, from the condition $X_{f}^{+}({\cal
T}_{c})=X_{f}^{-}({\cal T}_{c})=X_{c}$ for the point of the
individual collapse we get
$$
{\cal T}_{c}=-\beta_{d}(\chi_{s}-1)^{2}[1-{\cal
O}(\chi_{s}-1)+\cdots]
$$
$$
X_{c}=\sqrt{3(\chi_{s}-1+{\cal T}_{c})}
$$
where $\beta_{d}=3/2(d-1)$.

Determining further the distance of $\pm$ fronts to the collapse
point $\Delta_{\pm}=X_{f}^{\pm}-X_{c}$ and assuming that
$|\Delta_{\pm}|/X_{c}\ll 1$ we find from Eq. (47)
$$
\Delta_{\pm}=\pm\frac{\sqrt{6(d-1)({\cal T}-{\cal
T}_{c})}}{2X_{c}}(1-{\Delta_{\pm}/2X_{c}+\cdots}),
$$
whence it follows that
$$
\Delta_{+}/|\Delta_{-}|= 1-|\Delta_{\pm}|/X_{c}+\cdots\to 1
$$
as $|\Delta_{\pm}|/X_{c}\propto\sqrt{{\cal T}-{\cal
T}_{c}}/X_{c}^{2}\to 0$. Introducing now the difference coordinate
$\Delta=X-X_{c}$, from Eq.(47) we obtain expansion in powers of
$\Delta_{f}$ in the form $(d-1)({\cal T}- {\cal T}_{c}) = -2({\cal
T}-{\cal T}_{c})\Delta_{f}X_{c}+(2/3)\Delta_{f}^{2}X_{c}^{2}+{\cal
O}(X_{c}\Delta_{f}^{3}, \Delta_{f}^{4})+\varrho_{f}^{2}+\cdots$
whence it follows that at the final stage of individual collapse
$|\Delta_{f}|/X_{c}\ll 1$ each of the islands takes the shape of
an ellipse (2D) or ellipsoid of revolution (3D) with the center at
the collapse point $X_{c}$
\begin{eqnarray}
\left(\frac{\Delta_{f}}{\Delta_{f}^{m}}\right)^{2}+\left(\frac{\varrho_{f}}{\varrho_{f}^{m}}\right)^{2}=1,
\end{eqnarray}
the semi-axes of which contract by the laws
$$
\Delta_{f}^{m}= \frac{\sqrt{6(d-1)({\cal T}-{\cal
T}_{c})}}{2X_{c}},
$$
$$
\varrho_{f}^{m}=\sqrt{({\cal T}-{\cal T}_{c})(d-1)},
$$
and, consequently, the ellipse (ellipsoid) contracts {\it
self-similarly} up to the collapse point with the {\it
time-independent} aspect ratio
$$
{\cal A}=\varrho_{f}^{m}/\Delta_{f}^{m}=\sqrt{2/3}X_{c}\propto
\sqrt{\chi_{s}-1},
$$
so that ${\cal A}\to 0$ as $\chi_{s}\to 1$. According to Eq. (47),
during evolution from the fragmentation point ${\cal T}=0$ to the
collapse point ${\cal T}={\cal T}_{c}$ the island center almost
does not shift $(X_{\star}(0)-X_{c})/X_{c}\propto \chi_{s}-1\ll
1$, the concentration of $A$-particles in the island center
decreases by the law $s_{c}\propto ({\cal T}-{\cal T}_{c})$ and
the fraction of particles remaining in the island decreases by the
law ${\cal N}/{\cal N}_{0}\propto ({\cal T}-{\cal
T}_{c})^{(d+2)/2}/\sqrt{\chi_{s}-1}$. As an illustration, Fig.8
shows the evolution of the shape of 2D islands from the
"hyperbolic" ($|{\cal T}|\ll {|\cal T}_{c}|$) to the "elliptical"
stage (${\cal T}-{\cal T}_{c}\ll |{\cal T}_{c}|$) stage for
$\chi_{s}-1=0.01$. Below we shall demonstrate that, as well as in
the case of death of the single-centered island ($T_{c}>T_{s}$),
at the final stage of individual death ($T_{c}<T_{s}$) each of the
islands takes the shape of an ellipse (ellipsoid of revolution) at
any $\Lambda< \Lambda_{s}$, degenerating into a $d$-dimensional
sphere in the limit of autonomous collapse $\Lambda^{2/d}\ll 1$.

\begin{figure}
\includegraphics[width=1\columnwidth]{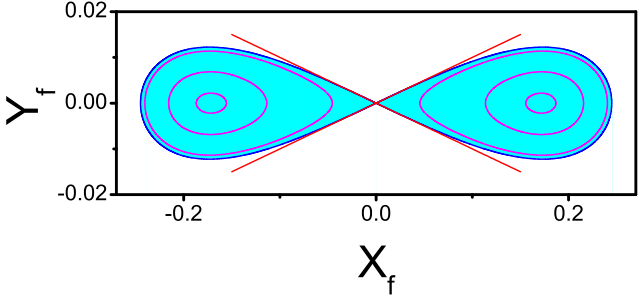}
\caption {Evolution of 2D island from the fragmentation $({\cal
T}=(T_{fr}-T)/T_{fr}=0)$ to the collapse $({\cal T}={\cal T}_{c})$
points calculated from Eq. (47) in the quasi-1D "string" limit
$\chi_{s}-1=0.01$ for the time moments ${\cal T}=0, -2\times
10^{-5}, -10^{-4}$ and $-1.41\times 10^{-4}$.}
\label{Fig8}
\end{figure}

\subsection{Final stage of the individual collapse of islands
($\Lambda< \Lambda_{s})$}

Let now, as before, $\Delta=X-X_{c}$, but ${\cal
T}=(T_{c}-T)/T_{c}$ where $T_{c}$ is the time moment of the
individual island collapse ($\Lambda< \Lambda_{s}$). Then in the
limit of small $\Delta^{2}/T_{c}\ll 1$, $\varrho^{2}/T_{c}\ll 1$
and ${\cal T}\ll 1$ from Eq. (20) we obtain expansion in powers of
$\Delta, \varrho$ and ${\cal T}$ in the form
\begin{eqnarray}
\frac{s-s_{c}}{1+s_{c}}= {\cal F}(\Delta)+{\cal
E}({\varrho})(1+{\cal F}(\Delta))
\end{eqnarray}
where concentration of $A$-particles at the collapse point
$\Delta, \varrho=0$ decreases by the law
$$
s_{c}={\cal T}(d-\chi_{c})/2+m_{c}{\cal T}^{2}+\cdots,
$$
$$
m_{c}=\frac{(d+2)}{8}(d-2\chi_{c})+\frac{\chi_{c}(1+3X_{c}^{2})}{16T_{c}},
$$
and expansions of the functions ${\cal F}(\Delta)$ and ${\cal
E}({\varrho})$ in powers of $\Delta$ and $\varrho$, respectively,
have the form
$$
{\cal F}(\Delta)=c_{1}\Delta + c_{2}\Delta^{2}+ c_{3}\Delta^{3}+
c_{4}\Delta^{4}+\cdots,
$$
and
$$
{\cal E}(\varrho)= -\varrho^{2}(1+{\cal
T}-\varrho^{2}/8T_{c}+\cdots)/4T_{c}+\dots
$$
where the coefficients
$$
c_{1}=\frac{\chi_{c}X_{c}{\cal T}}{2T_{c}}(1+{\cal O}({\cal
T})+\cdots),
$$
$$
c_{2}= - \frac{1-\chi_{c}+\omega{\cal T}+\cdots}{4T_{c}},
$$
$$
\omega=1+\chi_{c}(X_{c}^{2}/T_{c}-2),
$$
$$
c_{3}= - \frac{\chi_{c}X_{c}}{12T_{c}^{2}}(1+{\cal O}({\cal
T})+\cdots),
$$
$$
c_{4}=\frac{1+\chi_{c}(X_{c}^{2}/2T_{c}+1/6T_{c}-2)}{32T_{c}^{2}}(1+{\cal
O}({\cal T})+\cdots)
$$
and the notation is introduced
\begin{eqnarray}
0<\chi_{c}=\frac{(1-X_{c}^{2})}{2T_{c}}<1.
\end{eqnarray}
From Eq. (49) we conclude that at any $\Lambda < \Lambda_{s}$ at
the final collapse stage ${\cal T}\to 0$ the distribution of
particles in each of the islands takes the universal scaling form
\begin{eqnarray}
s = {\cal T}{\cal S}_{d,\Lambda}\left(\frac{|\Delta|}{{\cal
T}^{1/2}}, \frac{|\varrho|}{{\cal T}^{1/2}}\right),
\end{eqnarray}
whence, according to Eq. (49), it follows that in 1D systems in
the domain $X>=0$ the fronts $\Delta_{\pm}=X_{f}^{\pm}-X_{c}$
asymptotically converge symmetrically to the collapse point
$\Delta_{\pm}=0$ by the law
$$
\Delta_{\pm}=\pm\Delta_{f}^{m}=\pm\sqrt{2T_{c}{\cal T}}.
$$
Correspondingly, in agreement with Eq. (48), in 2D and 3D systems
each of the islands takes asymptotically the shape of an ellipse
(2D) or ellipsoid of revolution (3D) the semi-axes of which
contract by the laws
\begin{eqnarray}
\Delta_{f}^{m}=\sqrt{\frac{2T_{c}(d-\chi_{c}){\cal
T}}{(1-\chi_{c})}},
\end{eqnarray}
\begin{eqnarray}
\varrho_{f}^{m}=\sqrt{2T_{c}(d-\chi_{c}){\cal T}},
\end{eqnarray}
and, consequently, the ellipse (ellipsoid) contracts {\it
self-similarly} up to the collapse point with the time-independent
aspect ratio
$$
{\cal A}=\varrho_{f}^{m}/\Delta_{f}^{m}=\sqrt{1-\chi_{c}}.
$$
According to Eq. (22), in the limit $\Lambda\to \Lambda_{s}$ we
find from Eq. (50) $1-\chi_{c}= 2X_{c}^{2}/3+\cdots\to 0$ as
$T_{c}\to T_{s}$, whereas in the opposite limit
$\Lambda/\Lambda_{s}\ll 1$($T_{c}\ll T_{s}$) the value of
$\chi_{c}$ rapidly becomes exponentially small with a decrease in
$\Lambda$: $\chi_{c}\propto e^{-1/2T_{c}}/T_{c}\to 0$ as $T_{c}\to
0$. Thus, we conclude that in the limit $\Lambda\to \Lambda_{s}$
($\chi_{c}\to 1$), as expected, the island "inherits" the shape of
a quasi-1D "string" ${\cal A}(\chi_{c}\to 1)\to 0$, whereas, in
agreement with Eq. (28), in the opposite limit of autonomous death
$\Lambda\ll \Lambda_{s}$ ($\chi_{c}\ll 1$) the island contracts
self-similarly in the form of a $d$-dimensional sphere, ${\cal
A}(\chi_{c}\to 0)\to 1$.

In the general case for the trajectories of crossover to the
regime of self-similar collapse of $\pm$ fronts along the $X
(\varrho_{f}=0)$ axis we obtain from Eq. (49)
\begin{eqnarray}
\Delta_{\pm}=\pm\Delta_{f}^{m}(1\mp q{\cal T}^{1/2}+ g{\cal T} +
\cdots)
\end{eqnarray}
where
$$
q =
\frac{\chi_{c}X_{c}\sqrt{2(d-\chi_{c})/T_{c}}}{6(1-\chi_{c})^{3/2}}(1-\Gamma),
\quad \Gamma=\frac{3(1-\chi_{c})}{(d-\chi_{c})},
$$
and
$$
g = \frac{m_{c}}{d-\chi_{c}}- \frac{\omega}{2(1-\chi_{c})}+
(d-\chi_{c})\left[\frac{c_{4}T_{c}}{|c_{2}|(1-\chi_{c})}-1/4\right].
$$
Determining the domain of self-similar island collapse by the
condition ${\sf Max}||\Delta_{\pm}|/\Delta_{f}^{m}-1|< \epsilon\ll
1$ we find from Eq. (54) that in the limit of small $1-\chi_{c}\ll
1$ the boundary of this domain is determined by the dominant term
$q{\cal T}^{1/2}$
\begin{eqnarray}
{\cal T}_{\pm}\sim (\epsilon/q)^{2}\sim
\frac{6(1-\chi_{c})^{2}\epsilon^{2}}{(d-\chi_{c})(1-\Gamma)^{2}},
\end{eqnarray}
whereas in the opposite limit $\chi_{c}\sim
\exp(-1/2T_{c})/T_{c}\ll 1$ the boundary of this domain (in
agreement with the dynamics of autonomous collapse of the
$d$-dimensional sphere Eq. (28)) is determined by the dominant
term $g{\cal T}$
\begin{eqnarray}
{\cal T}_{\pm}\sim \epsilon/|g| \sim 4\epsilon.
\end{eqnarray}
In Fig. 9 are shown the dependencies ${\cal T}_{+}(T_{c})$ and
${\cal T}_{-}(T_{c})$ for $\epsilon=0.01$ which demonstrate the
key features of crossover to the regime of self-similar collapse
of $\pm$ fronts at $d=1,2$ and 3. It is seen that in 1D and 3D
systems the boundary of the self-similar collapse regime ${\sf
Max}||\Delta_{\pm}|/\Delta_{f}^{m}-1|=\epsilon$ is determined by
evolution of either the front $\Delta_{-}$($d=1, q<0$) or the
front $\Delta_{+}$($d=3, q>0$), whereas in the $2D$ case this
boundary is determined by evolution of the front $\Delta_{-}$ in
the range $0<\chi_{c}<1/2 (q<0)$ and by evolution of the front
$\Delta_{+}$ in the range $1/2<\chi_{c}<1 (q>0)$. We emphasize
that according to Eqs. (55) and (56) in the ranges $T_{c}< 0.1$
and $T_{c}> 0.3-0.4$ both of the fronts reach the boundary of the
self-similar collapse regime simultaneously. Moreover, the
detailed analysis shows that in 1D systems $\Delta_{+}>
|\Delta_{-}|$, in 3D systems $\Delta_{+}<|\Delta_{-}|$ and in 2D
systems $\Delta_{+}>|\Delta_{-}|$ in the range $0<\chi_{c}<1/2$,
whereas in the range $1/2<\chi_{c}<1$, where the individual
collapse is preceded by coalescence-fragmentation,
$\Delta_{+}<|\Delta_{-}|$. According to Eq. (49), in 2D and 3D
systems the "trajectory" of crossover to the regime of
self-similar contraction of minor semi-axis of the ellipse
(ellipsoid) has the form
$$
|\varrho_{f}(\Delta=0)|=\varrho_{f}^{m}(1+\phi{\cal T}+\cdots)
$$
where $\phi= m_{c}/(d-\chi_{c})-(d-\chi_{c})/8 -1/2$, whence it
follows that in fulfilling the requirement ${\sf
Max}||\Delta_{\pm}|/\Delta_{f}^{m}-1|< \epsilon$ the condition
$||\varrho_{f}(\Delta=0)|/\varrho_{f}^{m}-1|< \epsilon$ is
satisfied automatically in the entire range
$0<\Lambda<\Lambda_{s}$.

It is easy to see that, as it should be, in the collapse area of a
single-centered island, $\Lambda>\Lambda_{s}$, the Eq. (54 ) is
transformed into Eqs. (41), where $X_{c}=0,
\chi=\chi_{c}=1/2T_{c}, q=0, \mu=g, X_{f}^{m}=\Delta_{f}^{m},$ and
${\cal T}(\epsilon)\sim \epsilon/|\mu|$ in the whole range
$T_{c}>T_{s}$.

\begin{figure}
\includegraphics[width=1\columnwidth]{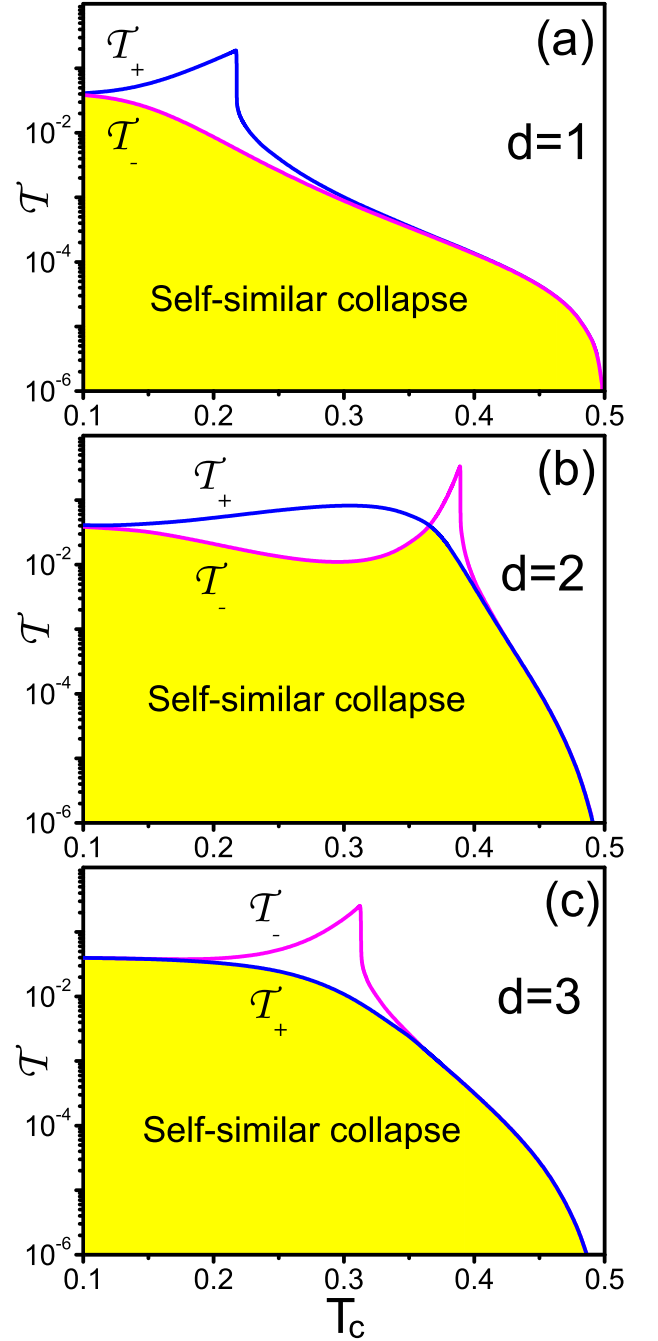}
\caption {Dependencies ${\cal T}_{\pm}(T_{c})$ calculated from Eq.
(21) at $\epsilon=0.01$ for 1D (a), 2D (b) and 3D (c) systems. The
areas of self-similar collapse are colored.}
\label{Fig9}
\end{figure}

\section{Evolution, delocalization and relocalization of the reaction front}

So far we have assumed that the reaction front is sharp enough so
that the front relative width remains negligibly small and, as a
consequence, the front moves quasistatically up to a narrow
vicinity of the island collapse point. In this section, we shall
reveal the conditions for this assumption realization.

In Refs. \cite{cor1},\cite{lee}, \cite{ben}, \cite{koza} it has
been established that at $d>d_{c}=2$ in the dimensional variables
the dependence of the quasistatic front width $w$ on the boundary
current density $J$ is described by the mean-field law,
\begin{eqnarray}
w_{{\sf MF}}\sim (D^{2}/kJ)^{1/3},
\end{eqnarray}
whereas in the 1D case in the diffusion-controlled limit the
quasistatic front width becomes $k$-independent and it is
determined by the fluctuation law
$$
w_{{\sf F}}\sim \sqrt{D/J}
$$
(at upper critical dimension $d_{c}=2$ in the mean-field law Eq.
(57) a logarithmic correction appears \cite{kra}, \cite{self7}).

As noted in the Introduction, in the case of autonomous evolution
of the $d$-dimensional spherical island, an exhaustive analysis of
the reaction front relative width evolution for the fluctuation,
the logarithmically modified, and the mean-field regimes was
presented \cite{self7}. It was demonstrated that in wide range of
parameters at a large enough initial island particle number the
front remains sharp up to a narrow vicinity of the island collapse
point, and therefore the whole picture of island evolution is
completely self-consisted. According to Ref. \cite{self7} in the
mean-field regime (i.e. for quasi-1D, quasi-2D and 3D systems)
regardless of the system dimension and the initial number of
island particles, evolution of the relative front width
$\eta=w/r_{f}$ is described by a universal law within which the
characteristic time of front delocalization at the final collapse
stage is determined unambiguously by the relative width of the
front at its turning point ${\cal T}_{Q}\sim \eta_{M}^{3/2}$.

From the analysis presented above it is clear that in the problem
of evolution of the island-sea-island system behavior of the front
width becomes much more complicated except for the limits of small
and large $\Lambda$, where the majority of particles die in the
regime of evolution of the $d$- dimensional sphere. Fortunately,
the detailed description of behavior of the front width $w({\bf
r}_{f},\Lambda,T)$ is not necessary. Indeed, just as in the case
of evolution of the $d$-dimensional sphere, we will be primarily
interested in revealing the parameter domain within which the
front delocalization occurs at the final (self-similar) collapse
stage where the front width grows unlimitedly as $T$ approaches
the collapse point $T_{c}$. Our second main aim will be to
determine the parameter domain within which front delocalization
occurs in a narrow vicinity of the points of coalescence $T_{cl}$
and fragmentation $T_{fr}$ where the front width increases
unlimitedly as the front approaches the system center.

To avoid unnecessary complications, we will consider evolution of
the front in the mean-field regime for quasi-1D, quasi-2D and 3D
systems. According to Eq. (57), in the units that we have accepted
, the mean-field front width reads
\begin{eqnarray}
w\sim 1/({\kappa J})^{1/3}
\end{eqnarray}
where the effective reaction constant $\kappa=kb_{0}\ell^{2}/D$
and the boundary current density $J=|\nabla s||_{\bf
r_{f}}=(\sqrt{(\partial_{X}
s)^{2}+(\partial_{\varrho}s)^{2}})|_{\bf r_{f}}$.

\subsection{Final stage of the individual collapse of islands $(\Lambda< \Lambda_{s})$}

At $\Lambda< \Lambda_{s}$ from Eq. (49) we find that on the final
(self-similar) collapse stage the relative width of the reaction
front along the $X$ axis increases by the law
\begin{eqnarray}
\eta_{\Delta}^{m}=w_{\Delta}/\Delta_{f}^{m}= \left(\frac{{\cal
T}_{\Delta}^{Q}}{{\cal T}}\right)^{2/3}
\end{eqnarray}
where the characteristic time of front delocalization is
\begin{eqnarray}
{\cal T}_{\Delta}^{Q}=\sqrt{\frac{(1-\chi_{c})}{2\kappa
T_{c}(d-\chi_{c})^{2}}}.
\end{eqnarray}
In the quasi-2D and 3D systems Eqs. (59) and (60) determine the
evolution of the relative front width along the major semi-axis of
the ellipse (ellipsoid), whereas the evolution of the relative
front width along the minor semi-axis of the ellipse (ellipsoid)
is determined by the law
\begin{eqnarray}
\eta_{\varrho}^{m}=w_{\varrho}/\varrho_{f}^{m}= \left(\frac{{\cal
T}_{\varrho}^{Q}}{{\cal T}}\right)^{2/3}
\end{eqnarray}
where the characteristic time of front delocalization is
\begin{eqnarray}
{\cal T}_{\varrho}^{Q}=1/\sqrt{2\kappa T_{c}(d-\chi_{c})^{2}}.
\end{eqnarray}
According to Eqs. (60) and (62), ${\cal T}_{\Delta}^{Q}/{\cal
T}_{\varrho}^{Q}=\sqrt{1-\chi_{c}}<1$, hence, since the front
width varies monotonically along its $\Delta_{m}\to\varrho_{m}$
contour, ${\cal T}_{\Delta}^{Q}$ and ${\cal T}_{\varrho}^{Q}$
determine, respectively, the low and upper bounds of the
characteristic time of front delocalization.

In the limit of autonomous collapse of the $d$-dimensional sphere
$\chi_{c}\ll 1$ ($T_{c}\ll T_{s}$) at $\sqrt{\kappa T_{c}}\gg 1$
we find
\begin{eqnarray}
{\cal T}_{\Delta}^{Q}\approx{\cal T}_{\varrho}^{Q}=1/d(2\kappa
T_{c})^{1/2}
\end{eqnarray}
whence, in combination with Eqs. (52) and (53), for the
characteristic radius of front delocalization it follows
\begin{eqnarray}
\Delta_{f}^{Q}\approx
\varrho_{f}^{Q}=\left(\frac{2T_{c}}{\kappa}\right)^{1/4}
\end{eqnarray}
In the opposite limit $1-\chi_{c}\ll 1$ (i.e. in the vicinity of
the threshold of single-centered island collapse
$(T_{s}-T_{c})/T_{s}\ll 1$) we have ${\cal T}_{\Delta}^{Q}\ll
{\cal T}_{\varrho}^{Q}$ and, consequently, in a drastic contrast
to the quasi-1D case, at $d>1$ the front delocalization occurs
along the minor semi-axis of the ellipse (ellipsoid), that is why
these systems should be considered separately.

\subsubsection{Quasi-1D systems}

Satisfying the condition of self-similar collapse ${\cal
T}_{Q}^{\Delta}\ll 1-\chi_{c}\ll 1$ we find from Eq.(60) that this
condition is complied at
$$
1-\chi_{c}\gg \kappa^{-1/3}
$$
whence, according to Eqs. (52) and (60), we derive
\begin{eqnarray}
\kappa^{-1/2}\ll {\cal T}_{\Delta}^{Q}\sim \frac{1}{\sqrt{\kappa
(1-\chi_{c})}}\ll \kappa^{-1/3},
\end{eqnarray}
$$
\kappa^{-1/4}\ll\Delta_{m}^{Q}\sim \frac{1}{(\kappa
(1-\chi_{c}))^{1/4}}\ll \kappa^{-1/6}.
$$
In the opposite limit $1-\chi_{c}\ll{\cal T}_{X}^{Q}\ll 1$, long
before reaching the regime of self-similar collapse (51), front
delocalization occurs in the self-similar regime (38) whence it
follows that
\begin{eqnarray}
\eta_{X}= w_{X}/X_{f}={\cal T}_{X}^{Q}/{\cal T}
\end{eqnarray}
where
\begin{eqnarray}
 {\cal T}_{X}^{Q}\sim \kappa^{-1/3}, \quad
X_{f}^{Q}\sim \kappa^{-1/6}.
\end{eqnarray}

\subsubsection{Quasi-2D and 3D systems}

According to Eqs. (60) and (62), at $d>1$ in the vicinity of the
threshold of single-centered island collapse $1-\chi_{c}\ll 1$ the
characteristic time of front delocalization is determined by the
value ${\cal T}_{\varrho}^{Q}\gg {\cal T}_{\Delta}^{Q}$, that is
why the condition for crossover to the self-similar collapse
regime (51) is the requirement ${\cal T}_{Q}^{\varrho}\ll
(1-\chi_{c})^{2}$ whence it follows
$$
1-\chi_{c}\gg \kappa^{-1/4}
$$
and so we have
\begin{eqnarray}
{\cal T}_{\varrho}^{Q}\sim
1/(\kappa(d-\chi_{c})^{2})^{1/2}\sim \kappa^{-1/2}
\end{eqnarray}
$$
\kappa^{-1/4}\ll \Delta_{m}^{Q}\ll\kappa^{-1/8}, \quad
\varrho_{m}^{Q}\sim \kappa^{-1/4}
$$
with the aspect ratio
$$
\kappa^{-1/8}\ll {\cal A}_{Q}\ll 1.
$$
In the opposite limit $1-\chi_{c}\ll \kappa^{-1/4}$, long before
reaching the regime of self-similar collapse, front delocalization
occurs at the stage of superellipse (superellipsoid) evolution
$1-\chi_{c}\ll {\cal T}\ll 1$ where, according to Eq.(40), the
relative front width along the major semi-axis of superellipse
grows by the law
\begin{eqnarray}
\eta_{X}= w_{X}/X_{f}^{m}=\left(\frac{{\cal T}_{X}^{Q}}{{\cal
T}}\right)^{1/2}
\end{eqnarray}
where $X_{f}^{m}=(6{\cal T}(d-\chi_{c})^{1/4}$ and ${\cal
T}_{X}^{Q}\sim \kappa^{-2/3}$. According to Eq. (61), as well as
in the case of $\Delta$-collapse, the relative front width of
superellipse (superellipsoid) along its minor semi-axis grows by
the law (61) and determines the characteristic time of front
delocalization ${\cal T}_{\varrho}^{Q}\sim
1/\kappa^{1/2}(d-\chi_{c})>{\cal T}_{X}^{Q}$. Assuming further
that $1-\chi_{c}\sim X_{c}^{2}\ll {\cal T}_{\varrho}^{Q}\sim
\kappa^{-1/2}$ (including the critical point $X_{c}=0,
\Lambda=\Lambda_{s})$ we find
\begin{eqnarray}
X_{f}^{Q}\sim
\kappa^{-1/8}, \quad \varrho_{f}^{Q}\sim \kappa^{-1/4}
\end{eqnarray}
with the aspect ratio
$$
{\cal A}_{Q}\sim  \kappa^{-1/8}
$$
so that ${\cal A}\to 0$ as $\kappa\to \infty$ (note that as it
should be $X_{c}\ll (X_{f}^{Q})^{2}\ll X_{f}^{Q}\ll 1)$.

\subsection{Final stage of collapse of the single-centered island ($\Lambda>\Lambda_{s}$)}

According to Eq. (41) at $\Lambda > \Lambda_{s}\geq
\Lambda_{\star}$ the island coalescence is completed by the
self-similar collapse of single-centered island in the system
center $X_{c}=0$. Repeating the calculations of the previous
section it is not difficult to demonstrate that all the results
obtained above for the individual collapse of islands remain valid
for the collapse of the single-centered island, with the only
difference being that at $\Lambda > \Lambda_{s}$
$\Delta_{f}^{m}=X_{f}^{m}$ and $\chi_{c}=\chi=1/2T_{c}$.

\subsection{Evolution of the reaction front in the vicinity of coalescence point $T_{cl}$}

In a drastic contrast to autonomous evolution of the
$d$-dimensional spherical island, where front delocalization
occurs only at the final collapse stage, at $\Lambda
> \Lambda_{\star}$ the relative front width along the $X$ axis starts increasing
unlimitedly as the leading point of the front $X_{f}^{-}$
approaches the system center (where $J\to 0$ and hence
$w_{X}\to\infty$ as $T\to T_{cl}$) and, as a consequence, the {\it
intermediate} front delocalization arises.

Assuming that $|{\cal T}|=|T_{cl}-T|/T_{cl}\ll {\sf
min}[(\chi_{M}-1), T_{cl}]$ we find from Eqs.(35)
\begin{eqnarray} \eta_{X}^{m}=w_{X}/X_{f}^{m}=({\cal
T}_{X}^{Q}/{\cal T})^{2/3}
\end{eqnarray} where the characteristic time of front
delocalization is
\begin{eqnarray}
{\cal T}_{X}^{Q}=\sqrt{\frac{(\chi_{s}-1)}{2\kappa
T_{cl}(\chi_{s}-d)^{2}}}
\end{eqnarray}
whence, according to Eq. (46), the characteristic "length" of
delocalization is
\begin{eqnarray}
X_{f}^{Q}\sim \left(\frac{2T_{cl}}{\kappa
(\chi_{s}-1)}\right)^{1/4}
\end{eqnarray}
Far from the critical point $\Lambda\gg \Lambda_{\star}$
($\chi_{M}=\chi_{s}/d=1/2dT_{cl}\gg 1$) from Eqs. (72) and (73) at
any $d$ it follows
\begin{eqnarray}
{\cal T}_{X}^{Q}\sim \kappa^{-1/2}, \quad X_{f}^{Q}\sim
(T_{cl}^{2}/\kappa)^{1/4}
\end{eqnarray}
whence satisfying condition ${\cal T}_{X}^{Q}\ll T_{cl}\ll 1$ we
conclude
$$
T_{cl}\gg \kappa^{-1/2}, \quad \kappa^{-1/2}\ll X_{f}^{Q}\ll
\kappa^{-1/4}.
$$
In the vicinity of the critical point $\Lambda\sim
\Lambda_{\star}$ behavior of the relative front width at $d=1$ and
$d>1$ differs qualitatively.

\subsubsection{Quasi-1D systems}

Assuming that $\chi_{M}-1=\chi_{s}-1\ll 1$ and satisfying the
condition of self-similar coalescence regime ${\cal T}_{X}^{Q}\ll
\chi_{M}-1$, from Eq. (72) we find that this requirement is
fulfilled at
$$
\chi_{s}-1\gg \kappa^{-1/3}
$$
whence, according to Eqs. (72) and (73), it follows
$$
\kappa^{-1/2}\ll {\cal T}_{X}^{Q}\ll \kappa^{-1/3}, \quad
\kappa^{-1/4}\ll X_{f}^{Q}\ll \kappa^{-1/6}.
$$
In the opposite limit $\chi_{M}-1\ll {\cal T}_{Q}^{X}\ll 1$ long
before reaching the self-similar coalescence regime (46) front
delocalization occurs in the self-similar regime (38) whence, as
well as in the case of collapse, Eqs. (66) and (67) follow.

\subsubsection{Quasi-2D and 3D systems at $T<T_{cl}$}

Satisfying the condition of the self-similar coalescence regime
${\cal T}_{X}^{Q}\ll \chi_{M}-1 \ll 1$ in the vicinity of the
critical point $\chi_{M}-1\ll 1$ , from Eq, (72) we find that this
requirement is fulfilled at
$$
\chi_{M}-1\gg \kappa^{-1/4}
$$
whence, according to Eqs.(72) and (73), it follows
$$
\kappa^{-1/2}\ll {\cal T}_{X}^{Q}\ll \kappa^{-1/4}, \quad
X_{f}^{Q}\sim \kappa^{-1/4}
$$
In the opposite limit $\chi_{M}-1\ll{\cal T}_{Q}^{X}\ll 1$
(including the critical point $\chi_{M}=1$), according to Eq.
(45), the front moves with a constant velocity and we derive
\begin{eqnarray}
\eta_{X}^{m}=w_{X}/X_{f}^{m}= \left(\frac{{\cal T}_{X}^{Q}}{{\cal
T}}\right)^{4/3}
\end{eqnarray}
where the characteristic time of front delocalization is
\begin{eqnarray}
{\cal T}_{X}^{Q}=\sqrt{2}\left(\frac{d-1}{\kappa
d}\right)^{1/4}\sim \kappa^{-1/4}
\end{eqnarray}
and, hence, the characteristic length of delocalization is
\begin{eqnarray}
X_{f}^{Q}=(d(d-1)\kappa)^{-1/4}\sim \kappa^{-1/4}.
\end{eqnarray}

\subsubsection{Quasi-2D and 3D systems at $T>T_{cl}$}

According to Eqs.(35), at ${\cal T}<0$ in the system center ${\bf
r}=0$ an excess of $A$-particles arises with an increase in which
an isthmus between the islands is formed limited by the sharp
reaction front. From Eqs.(35) and (46) it follows that under the
condition $|{\cal T}|\ll T_{cl}$ with growing $|{\cal T}|$ the
relative front width in the plane $X=0$, where radius of the
isthmus is minimal, contracts by the law
\begin{eqnarray}
\eta_{\varrho} = w_{\varrho}/\varrho_{f}^{m}= ({\cal
T}_{\varrho}^{Q}/|{\cal T}|)^{2/3}
\end{eqnarray}
where the characteristic time of formation of a two-centered
island limited by the sharp front is
\begin{eqnarray}
{\cal T}_{\varrho}^{Q} = \sqrt{\frac{\chi_{s}}{\kappa
(\chi_{s}-d)^{2}}}
\end{eqnarray}
and the corresponding characteristic radius of the isthmus is
\begin{eqnarray}
\varrho_{f}^{Q}=(2T_{cl}/\kappa)^{1/4}
\end{eqnarray}
In the limit of large $\chi_{M}\gg 1 (T_{cl}\ll T_{M})$ from
Eq.(79) we obtain ${\cal T}_{Q}^{\varrho}=\sqrt{2T_{cl}/\kappa}$
whence satisfying the requirement ${\cal T}_{Q}^{\varrho}\ll
T_{cl}$ we find $\kappa^{-1}\ll T_{cl}\ll T_{M}$ and therefore we
conclude
\begin{eqnarray}
\kappa^{-1}\ll {\cal T}_{Q}^{\varrho}\ll \kappa^{-1/2}, \quad
\kappa^{-1/2}\ll \varrho_{f}^{Q}\ll \kappa^{-1/4}.
\end{eqnarray}
In the opposite limit $\chi_{M}-1\ll 1$, by satisfying the
condition of the self-similar coalescence regime ${\cal
T}_{Q}^{\varrho}\ll \chi_{M}-1$ from Eq. (79) we find
$\chi_{M}-1\gg \kappa^{-1/4}$ whence it follows that
\begin{eqnarray}
\kappa^{-1/2}\ll {\cal T}_{\varrho}^{Q}\ll \kappa^{-1/4}, \quad
\varrho_{f}^{Q}\sim \kappa^{-1/4}
\end{eqnarray}
According to Eq. (35), at $\chi_{M}-1\ll |{\cal T}|\ll 1$ in the
system center ${\bf r=0}$ an excess of sea particles arises and,
consequently, in the vicinity of coalescence threshold
$\chi_{M}-1\ll \kappa^{-1/4}$ no isthmus between the islands is
formed. We conclude thus that formation of a two-centered
dumbbell-like island limited by the sharp front occurs only in the
domain $\chi_{M}-1\gg \kappa^{-1/4}$.

\subsection{Evolution of the reaction front in the vicinity of
fragmentation point $T_{fr}$}

\subsubsection{Quasi-2D and 3D systems at $T<T_{fr}$}

According to Eqs. (35), in the domain ${\cal
T}=(T_{fr}-T)/T_{fr}\ll 1-\chi_{M}$, as the excess of
$A$-particles in the system center decreases, the half-width
(radius) of the isthmus in the $X=0$ plane contracts by the law
(46), whence it follows that with growing $T$ the relative front
width increases by the law (78) where the characteristic time of
front delocalization and the corresponding characteristic
half-width (radius) of the isthmus are determined by Eqs. (78) and
(80) with $T_{fr}$$(1<\chi_{s}=T_{s}/T_{fr}<d)$ instead of
$T_{cl}$$(d<\chi_{s}=T_{s}/T_{cl}<\infty)$. As expected, in the
vicinity of fragmentation threshold $1-\chi_{M}\ll 1$ the front
delocalization occurs only in the domain $1-\chi_{M}\gg
\kappa^{-1/4}$ where at the coalescence stage the front
localization occurs
$$
(w\downarrow) \varrho_{f}\longleftarrow {\bf 0} \longrightarrow
-\varrho_{f} (w\downarrow), |{\cal T}|\uparrow, \quad (cl, {\bf
loc})
$$
$$
(w\uparrow) \varrho_{f}\longrightarrow {\bf 0} \longleftarrow
-\varrho_{f} (w\uparrow), {\cal T}\downarrow, \quad (fr, {\bf
deloc})
$$
Not too close to the fragmentation threshold up to the threshold
of single-centered island $\chi_{s}\sim 1$ we find from Eqs.(79)
and (80)
\begin{eqnarray}
{\cal T}_{\varrho}^{Q}\sim \kappa^{-1/2}, \quad
\varrho_{f}^{Q}\sim \kappa^{-1/4}.
\end{eqnarray}

\subsubsection{Quasi-2D and 3D systems at $T>T_{fr}$}

Just as during of coalescence the front delocalization occurs in
the vicinity of vertices of the hyperbola (hyperboloid)
$|X_{f}^{-}|$ moving towards each other,
$$
(w\uparrow) X_{f}^{-}\longrightarrow {\bf 0} \longleftarrow
-X_{f}^{-} (w\uparrow), {\cal T}\downarrow, \quad (cl, {\bf
deloc})
$$
at the final fragmentation stage the formation of the localized
front is completed in the vicinity of the vertices $|X_{f}^{-}|$
moving away from each other,
$$
(w\downarrow) X_{f}^{-}\longleftarrow {\bf 0} \longrightarrow
-X_{f}^{-} (w\downarrow), |{\cal T}|\uparrow, \quad (fr, {\bf
loc}).
$$
According to Eq. (35), at $|{\cal T}|\ll {\sf
min}[(\chi_{s}-1)^{2}, |\chi_{M}-1|, T_{fr}]$ with an increase in
$|{\cal T}|$ the relative front width in the vertices vicinity
contracts by the law
\begin{eqnarray} \eta_{X}^{m}=w_{X}/X_{f}^{m}=({\cal
T}_{X}^{Q}/|{\cal T}|)^{2/3}
\end{eqnarray}
where the characteristic time of front localization is
\begin{eqnarray}
{\cal T}_{X}^{Q}=\sqrt{\frac{(\chi_{s}-1)}{2\kappa
T_{fr}(\chi_{s}-d)^{2}}}
\end{eqnarray}
and the characteristic length of localization is
\begin{eqnarray}
X_{f}^{Q}\sim
\left(\frac{2T_{fr}}{\kappa(\chi_{s}-1)}\right)^{1/4}.
\end{eqnarray}

i) From Eqs. (72) and (85) we conclude that, as expected, in the
vicinity of the coalescence-fragmentation threshold $1-\chi_{M}\ll
1$ a remarkable symmetry takes place
$$
\eta_{X}(cl, {\cal T})\leftrightarrow \eta_{X}(fr, -{\cal T})
$$
and, as a consequence, in the domain $\kappa^{-1/4}\ll
1-\chi_{M}\ll 1$ the characteristic time of front localization
${\cal T}_{X}^{Q}(fr)\ll \kappa^{-1/4}$, whereas in the domain of
"elastic" front reflection $1-\chi_{M}\ll \kappa^{-1/4}$ we return
to Eqs. (76),(77).

ii) Much more nontrivial is the front localization in the limit of
a quasi-1D "string" $\chi_{s}-1\ll 1$. Assuming that ${\cal
T}_{X}^{Q}\ll |{\cal T}_{c}|\sim (\chi_{s}-1)^{2}$ we derive from
Eq. (85) that the front localization is completed long before the
collapse of the island in the domain $\chi_{s}-1\gg \kappa^{-1/3}$
where ${\cal T}_{X}^{Q}/|{\cal T}_{c}|\sim
(\kappa^{1/3}(\chi_{s}-1))^{-3/2}\ll 1, \kappa^{-2/3}\ll {\cal
T}_{X}^{Q}\ll |{\cal T}_{c}|$ and $X_{f}^{Q}\ll \kappa^{-1/6}\ll
X_{c}\sim \sqrt{\chi_{s}-1}.$ It is clear, however, that
fragmentation is completed by the formation of separated daughter
islands with the sharp front only under the condition that by the
moment ${\cal T}_{X}^{Q}$ of front localization along the $X$ axis
the relative front width $\eta_{\varrho}^{c}$ along the $\varrho$
axis, outgoing from the island center, remains small enough. From
Eqs. (35),(47) we obtain easily that at ${\cal T}_{X}^{Q}/|{\cal
T}_{c}|\ll 1$ the ratio $\eta_{\varrho}^{c}\sim [\kappa
(\varrho_{f}^{c})^{4}]^{-1/3}\sim
[\kappa(\chi_{s}-1)^{4}]^{-1/3}$, whence it follows that
fragmentation is completed by the formation of separated islands
with the sharp front under fulfilling the more rigid condition
$\chi_{s}-1\gg \kappa^{-1/4}$ (note that this condition correlates
with the condition of self-similar $\Delta$-collapse
$1-\chi_{c}\gg \kappa^{-1/4}$), where $\kappa^{-5/8}\ll {\cal
T}_{X}^{Q}\ll |{\cal T}_{c}|/\kappa^{1/8}$ and $X_{f}^{Q}\ll
\kappa^{-3/16}\ll X_{c}$. We conclude thus that in the vicinity of
the critical point $\chi_{s}-1\ll \kappa^{-1/4}$ the fragmentation
of the quasi-1D "string" is completed by loss of individuality
(mixing with the sea) of the daughter islands long before
occurrence of the self-similar collapse stage.

iii) In the domain of intermediate $\Lambda_{\star}<\Lambda<
\Lambda_{s}$ not too close to the threshold points
$\Lambda_{\star}$ and $\Lambda_{s}$ we find from Eqs. (85) and
(86)
\begin{eqnarray}
{\cal T}_{X}^{Q}\sim \kappa^{-1/2}, \quad X_{f}^{Q}\sim
\kappa^{-1/4}.
\end{eqnarray}
The presented analysis of the key features of evolution of the
reaction front shows that the revealed picture of island
coalescence, fragmentation and collapse is completely
self-consistent only in the limit when the effective reaction
constant $\kappa=kb_{0}\ell^{2}/D$ is large enough. In the next
section our aim will be to demonstrate that in the
diffusion-controlled annihilation regime the value of $\kappa$ is
large indeed in a wide range of parameters.

\section{Evolution of the front in the diffusion-controlled annihilation regime}

Extracting the parameter $\Lambda$ in $\kappa$ explicitly, we
obtain
$$
\kappa=\frac{ka_{0}\ell^{2}}{cD}=\frac{ka_{0}\ell^{2}}{\Lambda D
L^{d}}=\frac{ka_{0}h^{d}\ell^{2-d}}{\Lambda D}
$$
whence, substituting here the constant of diffusion-controlled
annihilation in the 3D medium $k=\varsigma D r_{a}$ where $r_{a}$
is the annihilation radius and $\varsigma=8\pi$, we find
\begin{eqnarray}
\kappa={\cal K}/\Lambda, \quad {\cal K}=\varsigma
r_{a}a_{0}h^{d}\ell^{2-d}
\end{eqnarray}
From Eq. (88) it follows that (i) at fixed values of $\Lambda,
a_{0}$ and $L=\ell/h$ a simultaneous increase in the initial size
of the island and the initial distance between the islands results
in a rapid growth of ${\cal K}\propto \ell^{2}$; (ii) at fixed
values of $\Lambda, a_{0}$ and $h$ (i.e. at fixed initial particle
number in the island) with growing $\ell$ the value of ${\cal K}$
increases in quasi-1D systems, {\it does not change} in quasi-2D
systems and decreases in 3D systems as a consequence of a decrease
in the initial sea density $\propto 1/\ell^{d}$:
\begin{eqnarray}
\nonumber
{\cal K}\sim \left\{\begin{array} {lcl} {\sf const.}
\ell, \quad
d=1,\\
{\sf const.}, \quad d=2,\\
{\sf const.}/\ell, \quad d=3.\\
\end{array}\right.
\end{eqnarray}
Taking for illustration the realistic values $r_{a}\sim
10^{-8}cm$, $a_{0}\sim 10^{23}cm^{-3}$, $h=0.1cm$ and $\ell=10cm$
we find from Eq. (88) ${\cal K}\sim 10^{16}$ for $d=1$, ${\cal
K}\sim 10^{14}$ for $d=2$ and ${\cal K}\sim 10^{12}$ for $d=3$.
Below we will use these values of ${\cal K}$ to estimate the
typical domains of front delocalization.

\subsection{Collapse}

\subsubsection{Autonomous and single-centered collapse of spherical islands}

According to Eq. (28), in the limit of small $1/L^{2}\ll
\Lambda^{2/d}\ll 1$ evolution of each of the islands occurs in the
autonomous regime in the shape of a $d$-dimensional sphere with
the centers at the points $X_{c}\sim\pm 1$, whereas in the limit
of large $\Lambda^{2/d}\gg 1$ evolution of the formed by
coalescence single-centered island occurs in the shape of a
$d$-dimensional sphere with the center at the point $X_{c}=0$. It
is remarkable that in both limits evolution of the radius of the
island $\rho_{f} (T)$ and particle distribution in it is described
by the universal scaling laws Eqs. (16), (28) \cite{self7} with
the $\Lambda$-dependent time of island collapse $T_{c}\sim
\Lambda^{2/d}$. Substituting $T_{c}$ into Eqs. (63) and (64), for
the characteristic time ${\cal T}_{Q}$ and radius $\rho_{f}^{Q}$
of front delocalization at the final collapse stage we obtain
$$
{\cal T}_{Q}\sim ({\cal K}\Lambda^{(2-d)/d})^{-1/2}, \quad
\rho_{f}^{Q}\sim (\Lambda^{(d+2)/d}/{\cal K})^{1/4}
$$
whence taking into account the accepted parameters it follows
\begin{eqnarray}
{\cal T}_{Q}\sim \left\{\begin{array} {lcl} 10^{-8}/\Lambda^{1/2},
\quad
d=1,\\
10^{-7}, \quad d=2,\\
10^{-6}\Lambda^{1/6}, \quad d=3.\\
\end{array}\right.
\end{eqnarray}
and
\begin{eqnarray}
\rho_{f}^{Q}\sim \left\{\begin{array} {lcl} 10^{-4}\Lambda^{3/4},
\quad
d=1,\\
10^{-7/2}\Lambda^{1/2}, \quad d=2,\\
10^{-3}\Lambda^{5/12}, \quad d=3.\\
\end{array}\right.
\end{eqnarray}
Since, with the parameters fixed by us, an increase in
$\Lambda\propto {\sf const.}/b_{0}$ corresponds to a decrease in
the initial sea density $b_{0}$, from Eq. (89) it follows that
with a decrease in the initial sea density the characteristic time
of front delocalization ${\cal T}_{Q}$ decreases at $d=1$, {\it
does not change} at $d=2$ and increases slowly at $d=3$ remaining
small in a wide range of $\Lambda$. We conclude thus that in a
wide range of small and large $\Lambda$ the front remains sharp up
to a narrow vicinity of the collapse point. According to Eq. (90),
at any $d$ the characteristic delocalization radius $\rho_{f}^{Q}$
increases with growing $\Lambda$, remaining relatively small in a
wide range of $\Lambda$. More revealing is the ratio $\Xi_{Q} =
\rho_{f}^{Q}/\rho_{f}^{M}$ of the delocalization radius to the
radius of maximal island expansion $\rho_{f}^{M}\sim
\Lambda^{1/d}$ at the front turning point
\begin{eqnarray}
\nonumber
\Xi_{Q}\sim \sqrt{{\cal T}_{Q}}\sim \left\{\begin{array}
{lcl} 10^{-4}/\Lambda^{1/4}, \quad
d=1,\\
10^{-7/2}, \quad d=2,\\
10^{-3}\Lambda^{1/12}, \quad d=3.\\
\end{array}\right.
\end{eqnarray}
Here it should be emphasized a remarkable fact that at $d=2$ this
ratio, as well as ${\cal T}_{Q}$ and the relative front width at
the turning point $\eta_{\rho}^{M}\sim {\cal T}_{Q}^{2/3}$
\cite{self7}), {\it does not depend} on the initial sea density.

\subsubsection{Collapse of islands in the vicinity of the critical point
$\Lambda\sim \Lambda_{s}$}

In the quasi-1D case with an accuracy to an order of magnitude in
the range $10^{-5}\ll 1-\chi_{c}\ll 1$ we have ${\cal T}_{Q}\ll
10^{-5}$ and $\Delta_{f}^{Q}, X_{f}^{Q}\ll 10^{-3}$, whereas at
the critical point vicinity $1-\chi_{c}\ll 10^{-5}$ we find ${\cal
T}_{Q}\sim 10^{-5}$ and $X_{f}^{Q}\sim 10^{-3}$.

In quasi-2D systems at $1-\chi_{c}\ll 1$ we have ${\cal T}_{Q}\sim
10^{-7}, \varrho_{f}^{Q}\sim 10^{-4}$ where in the domain
$1-\chi_{c}\ll 10^{-7}$ of the self-similar "superelliptical"
collapse the aspect ratio by the moment of front delocalization is
${\cal A}_{Q}\sim 10^{-2}$, whereas in the domain
$10^{-4}\ll1-\chi_{c}\ll 1$ of the self-similar "elliptical"
collapse the aspect ratio is ${\cal A}_{Q}\sim
\sqrt{1-\chi_{c}}\gg 10^{-2}$. In 3D systems, respectively, at
$1-\chi_{c}\ll 1$ we find ${\cal T}_{Q}\sim 10^{-6},
\varrho_{f}^{Q}\sim 10^{-3}$ where in the domain $1-\chi_{c}\ll
10^{-6}$ of the self-similar "superellipsoidal" collapse the
aspect ratio is ${\cal A}_{Q}\sim 10^{-3/2}$, whereas in the
domain $10^{-3}\ll1-\chi_{c}\ll 1$ of the self-similar
"ellipsoidal" collapse the aspect ratio is ${\cal A}_{Q}\gg
10^{-3/2}$. Thus, we conclude that although in the quasi-1D case
in the vicinity of the critical point the quantity ${\cal T}_{Q}$
passes through a relatively sharp local maximum, at all $d$ the
front remains sharp up to a narrow vicinity of the collapse point.

\subsection{Coalescence}

\subsubsection{Coalescence far away from the threshold $\Lambda\gg
\Lambda_{\star}$}

Far away from the coalescence threshold $\Lambda\gg 1 (T_{cl}\sim
1/\ln\Lambda\ll 1)$ according to Eqs.(74) we find that in the
range ${\cal K}\sim 10^{16}-10^{12}(d=1,2,3)$ at $1\ll
\sqrt{\Lambda}\ln\Lambda\ll 10^{8}-10^{6}$ the characteristic time
of front delocalization is ${\cal T}_{X}^{Q}\sim
(10^{-8}-10^{-6})\sqrt{\Lambda}$ and the characteristic length of
front delocalization $X_{f}^{Q}\sim
(10^{-4}-10^{-3}(\Lambda/\ln^{2}\Lambda)^{1/4}$. As mentioned, in
a qualitative contrast to the quasi-1D case, where the internal
sea area disappears after the delocalization of the front
$X_{f}^{-}$, in quasi-2D and 3D systems the coalescence is
completed by formation of a dumbbell-like island limited by the
localized front with a minimal isthmus radius $\varrho_{f}^{Q}$ in
the system center $X_{f}=0$. According to Eqs. (69) and (70), the
characteristic time of front localization ${\cal T}_{\varrho}^{Q}$
and the isthmus radius $\varrho_{f}^{Q}$ are determined by the
expressions
$$
\varrho_{f}^{Q}\sim \sqrt{{\cal T}_{\varrho}^{Q}}\sim
(10^{-4}-10^{-3})(\Lambda/\ln\Lambda)^{1/4}.
$$
Since in the limit of too large $\Lambda\to\infty$
($T_{cl}/T_{c}\to 0$) the majority of particles die during
evolution of the formed spherical island, we conclude that in a
wide range of $\Lambda\gg 1$ the front remains sharp up to a
narrow vicinity of the coalescence point.

\subsubsection{Coalescence in the vicinity of the threshold $\Lambda\sim
\Lambda_{\star}$}

In the quasi-1D case in the vicinity of the coalescence threshold
$10^{-5}\ll \chi_{M}-1\ll 1$ we find
$$
{\cal T}_{X}^{Q}\ll 10^{-5}, \quad X_{f}^{Q}\ll 10^{-3},
$$
whereas in the domain $\chi_{M}-1\ll 10^{-5}$ we obtain ${\cal
T}_{X}^{Q}\sim 10^{-5}, X_{f}^{Q}\sim 10^{-3}$. In quasi-2D and 3D
systems at $T<T_{cl}$ in the range $10^{-4}-10^{-3} (d=2,3)\ll
\chi_{M}-1\ll 1$ we find
$$
{\cal T}_{X}^{Q}\ll X_{f}^{Q}\sim 10^{-4}-10^{-3},
$$
whereas in the domain $\chi_{M}-1\ll 10^{-4}-10^{-3}$ we obtain $
{\cal T}_{X}^{Q}\sim X_{f}^{Q}\sim 10^{-4}-10^{-3}.$
Correspondingly, at the final stage of coalescence at
$10^{-4}-10^{-3}\ll \chi_{M}-1\ll 1$ we have
$$
{\cal T}_{\varrho}^{Q}\ll \varrho_{f}^{Q}\sim 10^{-4}-10^{-3},
$$
whereas in a narrow vicinity of the threshold $\chi_{M}-1\ll
10^{-4}-10^{-3}$, instead of formation of an isthmus, the
"elastic" reflection of the "relocalized" front is realized at the
fragmentation stage ${\cal T}_{X}^{Q}({\sf fr})\sim {\cal
T}_{X}^{Q}({\sf cl})$.

\subsection{Fragmentation}

Due to narrowness of the fragmentation range, in the entire
fragmentation domain $\kappa \sim {\cal K}$, that is why,
according to Eqs. (83) and (87), not too close to the threshold
points ${\cal T}_{\varrho}^{Q}\sim {\cal T}_{X}^{Q}\sim
10^{-7}-10^{-6}$ and $\varrho_{f}^{Q}\sim X_{f}^{Q}\sim
10^{-4}-10^{-3}$. Correspondingly, in a narrow vicinity of the
coalescence threshold $1-\chi_{M}\ll 10^{-4}-10^{-3}$ "elastic"
reflection of the front occurs without coalescence, whereas in a
narrow vicinity of the threshold of centers merging $\chi_{s}-1\ll
10^{-4}-10^{-3}$ (the limit of the quasi-1D "string") the
fragmentation is completed by disruption of the islands.

Summarizing, we conclude that during diffusion-controlled
evolution of the islands the reaction front remains sharp up to a
narrow vicinity of the coalescence, fragmentation and collapse
points and, consequently, the whole picture of island evolution is
self-consistent in a wide range of parameters. According to
Eq.(88), with an increase in the initial particle number in the
island and a corresponding increase in the initial distance
between the islands this statement only gets stronger. Moreover,
due to a weak power dependence of ${\cal T}_{Q}$ on ${\cal K}$
this statement remains valid at a significant decrease in the
reaction constant.

\section{Conclusion}

In this paper, we have presented a systematic analytical study of
diffusion-controlled evolution, coalescence, fragmentation and
collapse of two identical spatially separated $d$-dimensional
$A$-particle islands in the $B$-particle sea at propagation of the
sharp reaction front $A+B\to 0$. The obtained self-consistent
picture of evolution of the islands and front trajectories is
based on three central assumptions: (i) on the condition of local
conservation of the difference concentration $s({\bf r},t)$ which
follows from the "standard" requirement of equality of unlike
particles diffusivities; (ii) on the assumption that the relative
front width can be neglected during islands evolution which
follows from the remarkable property of effective dynamical
"repulsion" of unlike species and (iii) on the quasi-static
approximation (QSA) which allowed obtaining a self-consistent
picture of front width evolution and revealing a domain of its
applicability parameters. The main results can be formulated as
follows:

1) It has been established that if the initial distance between
the centers of the islands $2\ell$ and the initial ratio of
concentrations island/sea $c=a_{0}/b_{0}$ are relatively large,
the evolution of the island-sea-island system is determined
unambiguously by the dimensionless parameter
$$
\Lambda={\cal N}_{0}/{\cal N}_{\Omega},
$$
where ${\cal N}_{0}$ is the initial particle number in the island
and ${\cal N}_{\Omega}$ is the initial number of sea particles in
the volume $\Omega= (2\ell)^{d}$.

2) It has been shown that there is a threshold value
$$
\Lambda_{\star}(d)=(\pi e/2d)^{d/2}/2,
$$
below which the islands die individually and above which island
coalescence occurs.

3) It has been established that regardless of $d$ the centers of
each of the islands move towards each other along the universal
trajectory, merging in a united center at the critical value
$$
\Lambda_{s}(d)=(\sqrt{e}/2)(\pi/2)^{d/2}.
$$
In 1D systems $\Lambda_{s}=\Lambda_{\star}$, that is why at
$\Lambda<\Lambda_{s}$ each of the islands dies individually,
whereas at $\Lambda > \Lambda_{s}$ coalescence is completed by the
collapse of the single-centered island in the system center. In 2D
and 3D systems in the range $\Lambda_{\star}<\Lambda <
\Lambda_{s}$ the coalescence is accompanied by the subsequent
fragmentation (division) of the two-centered island and is
completed by the individual collapse of each of the islands.

4) It has been demonstrated that in the limit of small
$\Lambda^{2/d}\ll 1$ the evolution of each of the islands-partners
occurs autonomously in the shape of a $d$-dimensional sphere with
the unshifted center. In the limit of large $\Lambda^{2/d}\gg 1$
the evolution of the island formed by coalescence occurs in the
shape of a $d$-dimensional sphere with the center in the system
center. In both of the limits the expansion-contraction-collapse
of the island is described by the {\it universal} scaling law.

5) It has been established that at any $d$ and $\Lambda$ the
evolution of the island in the vicinity of the collapse point
acquires a {\it self-similar} character. It has been shown that in
1D systems at $\Lambda\neq \Lambda_{s}$ "radii" of the islands
$\Delta_{\pm}$ contract "synchronously" to the collapse point
$X_{c}(\Lambda)$, whereas at the very critical point
$\Lambda_{s}=\Lambda_{\star}$ both of the islands die
simultaneously with the "internal" sea area. In 2D and 3D systems
at $\Lambda\neq \Lambda_{s}$ the island collapse occurs in the
shape of an ellipse (ellipsoid of revolution) with the constant
aspect ratio ${\cal A}(\Lambda)$ which contracts unlimitedly as
$\Lambda$ approaches the critical point $\Lambda_{s}$ both from
above and below (the limit of the "quasi-1D string"). At the very
critical point of centers merging $\Lambda_{s}$ the island
collapse occurs in the shape of a superellipse (superellipsoid of
revolution) with the aspect ratio ${\cal A}({\cal T})$ which
contracts unlimitedly with time while approaching the collapse
point $T_{c}$.

6) The laws of islands evolution in the vicinity of the starting
points of coalescence $T_{cl}(\Lambda)$ and fragmentation
$T_{fr}(\Lambda)$ have been revealed. It has been demonstrated
that in 2D and 3D systems the front takes the shape of a hyperbola
(hyperboloid of revolution) in the vicinity of the system center.
At $T<T_{cl}$ the vertices of the hyperbola (hyperboloid) move
towards each other forming at $T>T_{cl}$ a two-centered island
with an increasing isthmus radius in the system center. In the
vicinity of the fragmentation point this process occurs in a
reverse order. It has been shown that at the threshold point
$\Lambda_{\star}$ the "elastic" reflection of the front occurs in
the system center with an abrupt reversal of its velocity sign,
and a compact description of the island fragmentation-collapse in
the limit of the "quasi-1D string" $\Lambda\to \Lambda_{s}$ has
been found.

7) Within the QSA the self-consistent power laws of evolution of
the relative front width in the vicinity of coalescence,
fragmentation and collapse points have been revealed for quasi-1D,
quasi-2D and 3D systems. The characteristic times of front
delocalization and relocalization have been obtained depending on
the defining parameters of the problem. It has been shown that in
the diffusion-controlled annihilation regime the front remains
sharp up to a narrow vicinity of coalescence, fragmentation and
collapse points and, consequently, the whole picture is {\it
self-consistent} in a wide range of parameters.

As we have mentioned, because of mirror symmetry, this model
simultaneously describes the evolution of the $d$-dimensional
$A$-particle island in a semi-infinite $B$-particle sea with a
reflecting $(d-1)$-dimensional "wall". It should be emphasized,
however, that as well as in Ref.\cite{self7}, here the evolution
of islands has been considered at equal species diffusivities.
Although we believe that the regularities discovered reflect the
key features of islands evolution, the study of the much more
complicated problem for unequal species diffusivities remains a
challenging problem for the future. Moreover, we hope that the
future extensive numerical calculations together with the
corresponding experimental data will enable revealing a
comprehensive picture of evolution of the front during its
delocalization.

In conclusion we note that the mechanisms and regularities of
coalescence and collapse of two identical spatially separated
objects (liquid drops, biological cells, two-dimensional islands,
black holes, neutron stars etc.) in a foreign medium draw
increased interdisciplinary interest in a wide range of
applications from astrophysics, biophysics, and hydrodynamics to
condensed matter physics, chemical physics and materials science
\cite{man} - \cite{Reis}. Depending on nature of objects and
mechanisms of direct or indirect interaction with the medium, the
scenarios of coalescence and collapse, in spite of some common
features, demonstrate a rich diversity. We hope that the results
obtained in the present work represent one of the most detailed
scenarios of the coalescence, fragmentation and collapse
development, the basic features of which may turn out to be
universal in a wide spectrum of reaction-diffusion systems.

\begin{acknowledgments}

The research is carried out within the state task of ISSP RAS.

\end{acknowledgments}


\begin{references}
\bibitem{kbr} P.L. Krapivsky, E. Ben-Naim and S. Redner, {\it A
Kinetic View of Statistical Physics} (Cambridge University Press,
Cambridge, 2010)
\bibitem{tau} U.C. Tauber, M.Howard, and B.P. Vollmayr-Lee,
J.Phys. A {\bf 38}, R79 (2005)
\bibitem{rev} D. ben Avraham and S. Havlin, {\it Diffusion and Reactions in
Fractals and Disodered Systems} (Cambridge University Press,
Cambridge, 2000)
\bibitem{mat} D.C. Mattis and M.L. Glasser, Rev.Mod.Phys. {\bf
70}, 979 (1998)
\bibitem{cho} B. Chopard and M. Droz, {\it Cellular
automata modelling of physical systems} (Cambridge University
Press, Cambridge, 1998)
\bibitem{cot} E. Kotomin and V. Kuzovkov, {\it
Modern Aspects of Diffusion Controlled Reactions: Cooperative
Phenomena in Bimolecular Processes} (Elsevier, Amsterdam, 1996).
\bibitem{but1} L.V. Butov, A.C. Gossard and D.S. Chemla, Nature, {\bf 418 }, 751 (2002)
\bibitem{snoke} D. Snoke, S. Denev, Y. Liu, L. Pfeiffer and K.
West, Nature, {\bf 418 }, 754 (2002)
\bibitem{but2} Sen Yang, L.V. Butov, L.S. Levitov, B.D. Simons and
A.C. Gossard, Phys. Rev. B {\bf 81}, 115320 (2010)
\bibitem{ant} T. Antal, M. Droz, J. Magnin, and Z. Racz, Phys.
Rev. Lett., {\bf 83}, 2880 (1999).
\bibitem{rz} Z. Racz, Physica A, {\bf 274}, 50, (1999)
\bibitem{tho} S. Thomas, I. Lagzi, F. Molnar, Jr., and Z. Racz, Phys.
Rev. Lett., {\bf 110}, 078303 (2013).
\bibitem{br} F.Brau, G.Schuszter, and A.De Wit, Phys.Rev.Lett., {\bf 118}, 134101 (2017).
\bibitem{wit2} V.Loodts, P.M.J. Trevelyan, L. Rongy, and A.De
Wit, Phys.Rev.E, {\bf 94}, 043115 (2016).
\bibitem{must} I. Mastromatteo, B. Toth, and J.P. Bouchaud,
Phys.Rev.Lett., {\bf 113}, 268701 (2014).
\bibitem{gal} L. Galfi and Z. Racz, Phys. Rev. A {\bf 38}, 3151 (1988).
\bibitem{cor1} S. Cornell and M. Droz, Phys. Rev. Lett. {\bf 70}, 3824 (1993).
\bibitem{lee} B.P. Lee and J. Cardy, Phys. Rev. E  {\bf 50}, R3287 (1994).
\bibitem{cor2} S.J. Cornell, Phys.Rev.Lett., {\bf 75}, 2250 (1995).
\bibitem{kra} P.L. Krapivsky, Phys. Rev. E {\bf 51}, 4774 (1995).
\bibitem{bar} G.T. Barkema, M.J. Howard and J.L. Cardy, Phys. Rev. E {\bf 53}, R2017 (1996).
\bibitem{ben} E. Ben-Naim and S. Redner, J. Phys. A {\bf 25}, L575 (1992).
\bibitem{koza} Z. Koza, J. Stat. Phys. {\bf 85}, 179 (1996).
\bibitem{yus} S.B. Yuste, L. Acedo, and K. Lindenberg, Phys.Rev.E {\bf 69}, 036126 (2004).
\bibitem{sok} D. Froemberg and I.M. Sokolov, Phys.Rev.Lett. {\bf 100}, 108304 (2008).
\bibitem{kzt} Z. Koza and H.Taitelbaum, Phys.Rev.E {\bf 56}, 6387 (1997).
\bibitem{hec} I. Hecht, Y. Moran, and H. Taitelbaum, {\bf 73}, 051109 (2006).
\bibitem{rkz} I. Bena, M. Droz, K. Martens, and Z. Racz, J.Phys.:
Condens. Matter {\bf 19}, 065103 (2007).
\bibitem{self1} B.M. Shipilevsky, Phys. Rev. E {\bf 67}, 060101(R)(2003).
\bibitem{self2} B.M. Shipilevsky, Phys. Rev. E {\bf 70}, 032102(2004).
\bibitem{self3} B.M. Shipilevsky, Phys. Rev. E {\bf 77}, 030101(R)(2008).
\bibitem{kis} S. Kisilevich, M. Sinder, J. Pelleg, and V.
Sokolovsky, Phys. Rev. E {\bf 77}, 046103 (2008).
\bibitem{self4} B.M. Shipilevsky, Phys. Rev. E {\bf 79}, 061114 (2009).
\bibitem{self5} B.M. Shipilevsky, Phys. Rev. E {\bf 82}, 011119 (2010).
\bibitem{hay} C.P. Haynes, R. Voituriez, and O. Benichou, J. Phys. A {\bf 45}, 415001 (2012).
\bibitem{self6} B.M. Shipilevsky, Phys. Rev. E {\bf 88 }, 012133 (2013).
\bibitem{self7} B.M. Shipilevsky, Phys. Rev. E {\bf 95 }, 062137 (2017).
\bibitem{fial} M. Fialkowsky, A. Bitner, and B.A. Grzybovsky,
Phys.Rev.Lett. {\bf 94}, 018303 (2005)
\bibitem{lag} I. Lagzi, P.Papai, and Z.Racz, Chem.Phys.Lett. {\bf
433}, 286 (2007).
\bibitem{wit} N. Withers, Nat.Chem. {\bf 2}, 160 (2010).
\bibitem{man} A. Abi Mansour and M. Al-Ghoul, Phys.Rev. E {\bf
89}, 033303 (2014).
\bibitem{but3} L.V. Butov, L.S. Levitov, A.V. Mintsev, B.D.
Simons, A.C. Gossard, and D.S. Chemla, Phys.Rev.Lett. {\bf 92},
117404 (2004).
\bibitem{Gar} M. Garzon, L.J. Gray, and J.A. Sethian, Phys. Rev. E
{\bf 97}, 033112 (2018)
\bibitem{Des} T. Dessup, C. Coste, and M. Saint Jean, Phys. Rev.E
{\bf 95}, 012206 (2017)
\bibitem{Zha} J. Zhang, L. Chen, and M.-J. Ni, Phys.Rev. Fluids
{\bf 4}, 043604 (2019).
\bibitem{Shur} N.S. Shuravin, P.V. Dolganov, and V.K. Dolganov,
Phys.Rev. E {\bf 99}, 062702 (2019)
\bibitem{Peru} S. Perumanath, M.K. Borg, M.V. Chybynsky, J.E.
Sprittles, and J.M. Reese, Phys. Rev. Lett. {\bf 122}, 104501
(2019).
\bibitem{Car} C.M. Caragine, S.C. Haley, and A. Zidovska, Phys.
Rev. Lett. {\bf 121}, 148101 (2018)
\bibitem{Gup} A. Gupta, B. Krishnan, A.B. Nielsen, and E.
Schnetter, Phys.Rev. D {\bf 97}, 084028 (2018).
\bibitem{Reis} C. Reisswig, C.D. Ott, E. Abdikamalov, R. Haas, P.
Mosta, and E. Schnetter, Phys.Rev. Lett. {\bf 111}, 151101 (2013).

\end{references}
\end{document}